\documentclass[%
    reprint,
    aps,
    superscriptaddress,
    showpacs,
    prb,
    floatfix,
    bibnotes,
    amssymb,
    amsfonts,
]{revtex4-2}

\usepackage[utf8]{inputenc}
\usepackage[T1]{fontenc}
\usepackage{amsmath,amsthm,mathtools,mathrsfs,physics}
\usepackage{graphicx}
\usepackage{subfig}
\usepackage{hyperref}
\usepackage{color}
\usepackage{siunitx}

\graphicspath{{img/}}

\DeclareSIUnit\angstrom{\text{Å}}
\DeclareSIUnit\rydberg{Ry}

\begin{document}

\title{Quasiparticle Fermi surfaces of niobium and niobium-titanium alloys at high pressure}
\author{D. Jones}
\affiliation{Theoretische Physik III, Center for Electronic Correlations and Magnetism, Institute of Physics, University of Augsburg, 86135 Augsburg, Germany}
\affiliation{Augsburg Center for Innovative Technologies (ACIT), University of Augsburg, 86135 Augsburg, Germany}
\author{A. \"Ostlin}
\affiliation{Theoretische Physik III, Center for Electronic Correlations and Magnetism, Institute of Physics, University of Augsburg, 86135 Augsburg, Germany}
\author{A. Chmeruk}
\affiliation{Theoretische Physik III, Center for Electronic Correlations and Magnetism, Institute of Physics, University of Augsburg, 86135 Augsburg, Germany}
\author{F. Beiu\c seanu}
\affiliation{Faculty of Science, University of Oradea, 410087 Oradea, Romania}
\author{U. Eckern}
\affiliation{Theoretische Physik II, Institute of Physics, University of Augsburg, 86135 Augsburg, Germany}
\author{L. Vitos}
\affiliation{Applied Materials Physics, Department of Materials Science and Engineering, Royal Institute of Technology, Brinellv\"agen 23, 10044, Stockholm, Sweden}
\author{L. Chioncel}
\affiliation{Theoretische Physik III, Center for Electronic Correlations and Magnetism, Institute of Physics, University of Augsburg, 86135 Augsburg, Germany}
\affiliation{Augsburg Center for Innovative Technologies (ACIT), University of Augsburg, 86135 Augsburg, Germany}

% \date{\today}

\begin{abstract}
The electronic structure of pure niobium and the niobium-titanium alloy Nb$_{0.44}$Ti$_{0.56}$ in the \textit{bcc}-phase at pressures up to \SI{250}{\giga\pascal} is investigated, to reveal possible factors conducing toward the robust superconductivity reported for Ti-doped niobium upon a considerable volume reduction.
We model the structural disorder using the coherent potential approximation, and the electronic correlations are taken into account using dynamical mean-field theory. 
At high pressure, a significant change in the topology of the Fermi surface is observed, while electronic correlations weaken with increasing pressure.
Thus, the normal state of Nb$_{0.44}$Ti$_{0.56}$ is found to be a Fermi liquid with a well-defined Fermi surface, and well-defined quasiparticles near it. 
The systematic study of the impact of disorder upon the Fermi surface  at such ultra high pressures allows notable insights into the nature of the electronic states near the Fermi level, i.e., within the energy scale relevant for superconducting pairing. Furthermore, our results clearly indicate the necessity of further experimental Fermi surface explorations. 
\end{abstract}

\maketitle

\section{Introduction}
Pressure is an important external parameter that can be employed to alter the structure and the chemical bonding of a given material, and hence its physical, particularly its electronic properties. Using this tool in the regime of high pressures, significant modifications of material properties can be obtained, an example being the search for potential superconductors with high critical superconducting temperatures ($T_c$).

The understanding of the occurrence and robustness of superconductivity at high pressures has benefited substantially from the availability of structure predictions based on 
density functional theory (DFT)
electronic structure methods.
An example is the family of Nb-Ti alloys~\cite{gu.li.19,zh.ga.20,hu.gu.20}.
The superconductivity in this material is described using the Bardeen–Cooper–Schrieffer (BCS) phonon-mediated theory~\cite{ba.co.57,ba.co.57b}
in which the electron–phonon coupling parameter ($\lambda$) is considered as an indicator of the strength of potential superconductivity. 
A high pressure structural model was obtained using DFT methods including geometry optimization~\cite{sa.sa.96,ba.gi.01} and linear response computation of the $\lambda$~\cite{gius.17} parameter
at pressures up to \SI{250}{\giga\pascal}. 
The density of states at the Fermi level and the effective screened Coulomb constant $\mu^*$~\cite{mo.an.62} were also analyzed~\cite{zh.ga.20}.  
Employing the McMillan-Allen-Dynes~\cite{mcmi.68,al.dy.75} equation, $T_c$ was estimated to be located between \SIrange{15}{20}{\kelvin}, depending on the applied pressure, which agrees well with the reported experimental values, $T_c \simeq \SI{19}{\kelvin}$~\cite{gu.li.19}.

In spite of all these phenomenological studies, the nature of the robust superconducting phase at high pressures still requires further detailed investigation. 
One suggestion ~\cite{hu.gu.20} attributes the persistence of superconductivity at high pressure in Nb-Ti alloys to the almost constant density of states of e$_g$ orbitals at the Fermi level, while the partial density of states of t$_{2g}$ orbitals decreases continuously with increasing pressure. 
A similar behavior was reported in some multi-component high-entropy alloys, doping the  Nb-Ti materials~\cite{hu.gu.20} with further impurities.  

For the Nb$_{0.44}$Ti$_{0.56}$ alloy X-ray diffraction experiments showed that the \textit{bcc}-structure is retained through the whole range of pressures up to \SI{250}{\giga\pascal}. Disorder has been modeled by alternating atomic positions of Nb and Ti~\cite{zh.ga.20} with uniform occupation. Namely, the atoms occupy the $1a$ and $1b$ Wyckoff positions of the $Pm\overline{3}m$ CsCl lattice.
In a recent study~\cite{cu.na.24} state-of-the-art computations of the phonon spectra and the superconducting gap of Nb$_{0.5}$Ti$_{0.5}$ have been presented. Effects like anharmonic lattice dynamics and energy-dependent Coulomb interactions have been included, nevertheless the calculation largely overestimates $T_c$ compared to experiments. 
It was shown that, using a simple model based on Boltzmann-averaged supercells, the discrepancy can be qualitatively
understood in terms of lattice disorder. 

The results we present below use recent methodological developments beyond current standards, combining self-consistently disorder and correlation effects.
In particular, we do not make any assumption on the nature of the Ti distribution in the Nb \textit{bcc}-structure. 
Instead, we use the Coherent Potential Approximation (CPA)~\cite{sove.67,ve.ki.68,el.kr.74} and study the electronic structure of the disordered alloy including correlation effects captured by dynamical mean-field theory (DMFT)~\cite{ge.ko.96,ko.vo.04,held.07}, with specific tools borrowed from  DFT~\cite{os.vi.17,os.vi.18}. 
This allows us to perform a detailed analysis of the evolution of the Fermi surface (FS) of Nb and Nb$_{0.44}$Ti$_{0.56}$ at such high pressures. 
Possible explanations for the prediction of robust superconductivity at high pressures may also involve the existence of nesting (the coexistence of steep and flat bands in momentum space) in the first Brillouin zone.  
Quite unexpectedly, we find a dramatic evolution of the Fermi surface with increasing pressure (for pure Nb), and also in the presence of electronic correlations for the Nb$_{0.44}$Ti$_{0.56}$ alloy that to the best of our knowledge has not yet been reported in the literature.  

The paper is organized as follows: in Sec.~\ref{sec:methods} we provide a brief description of the computational methods within the DFT framework using pseudo-potential and muffin-tin based approaches; the computational details are discussed in Sec.~\ref{sec:comp_details}.
The analysis of the electronic structure of niobium and of the niobium-titanium alloy are presented in Sec.~\ref{sec:nb-ti}, including a discussion of electronic correlation effects. Finally, the Fermi surfaces results are presented in Sec.~\ref{sec:FS-results}. We conclude in Sec.~\ref{sec:summary}.

\section{Methods}
\label{sec:methods}
Density functional theory methods incorporating dynamical electronic correlations have proven essential in modeling physical properties of materials containing narrow-band electronic states~\cite{ko.vo.04,ko.sa.06,held.07}. 
Green's function based DFT-methods are very convenient when both the effects of dynamic electronic correlation and disorder are present, since the one-particle Green's function is computed directly and can be averaged according to the various disorder realizations~\cite{mi.eb.17,os.vi.17,os.vi.18,os.zh.20}. 

Our charge and self-energy self-consistent electronic structure method is using the exact muffin-tin orbitals (EMTO) basis set~\cite{an.je.94.2,vi.sk.00,vito.01} and extends the local density approximation (LDA) by including electronic correlations. 
Disorder is modeled using CPA~\cite{sove.67,ve.ki.68} which allows computations for any fractional concentration of disorder realizations in multi-component alloys. 
The scattering path operator is the central quantity in the multiple scattering formulations of the electronic structure. It simultaneously allows (i) the computation of the real space charge densities through charge self-consistency, and (ii) the inclusion of local dynamical correlation effects into the Green's function~\cite{os.zh.20,os.ch.21}.
The impurity solver produces the self-energy $\Sigma_\sigma(E)$ employing the method of spin-polarized $T$-matrix fluctuation exchange~\cite{li.ka.97,po.ka.06,ir.ka.08}; it corrects the starting DFT Green's function as discussed in previous publications~\cite{ch.vi.03,os.vi.17,os.vi.18}. 

The interaction between electrons is described considering the multi-orbital on-site interaction term:
\begin{equation}
    H_U = \frac{1}{2} \sum_{i{m,\sigma}} 
    U_{mm^{\prime} m^{\prime\prime}m^{\prime\prime\prime}} 
    c^\dagger_{im\sigma} 
    c^\dagger_{im^\prime \sigma^\prime} 
    c_{im^{\prime\prime\prime}\sigma^{\prime}}
    c_{im^{\prime\prime}\sigma}    
\end{equation}
Here $c_{im\sigma} (c^\dagger_{im \sigma})$ destroys (creates) an electron with spin $\sigma$ on orbital $m$ at the site $i$. 
The Coulomb matrix elements $U_{mm^{\prime}m^{\prime\prime}m^{\prime\prime\prime}}$ are parameterized in terms of the average local Coulomb $U$ and exchange $J$ parameters~\cite{ko.sa.06}. 
In principle, the dynamical electron-electron interaction matrix elements can be computed~\cite{ar.im.04}, however, substantial variations exist depending on the considered local orbitals~\cite{mi.ar.08}. 
We choose $U$ in the range from \SIrange{0}{5}{\eV}, and $J = \SI{0.6}{\eV}$.
We have checked that the results essentially do not change when increasing $J$ up to $\SI{0.9}{\eV}$.

To eliminate double counting of the interactions already included
in the exchange-correlation functional of the LDA, the self-energy $\Sigma_\sigma(E)$ is replaced by $\Sigma_\sigma(E)-\Sigma_\sigma(0)$ in all equations of the LDA+DMFT scheme. 
Throughout this paper finite temperatures are only considered for the electronic
subsystem, where $T$ enters in the Matsubara frequencies $\omega_n = (2n + 1)\pi T$, with $n=0, \ \pm 1, \ \pm 2, \ \dots \ $.
We perform total energy calculations using the 
LDA+DMFT functional~\cite{ko.sa.06} with the converged charge density and local Green’s function. 
In addition to the standard LDA total energy, the Kohn-Sham band energy correction due to the DMFT is added, together with the trace of the matrix product between the DMFT self-energy and the local Green's function ($\Sigma G$), the so-called Galitskii-Migdal contribution.

\section{Computational details}
\label{sec:comp_details}
The space group corresponding to the $\beta$-phase of Nb is $Im\overline{3}m$ with atoms occupying the Wyckoff $2a$ position. The pressure dependent lattice parameters for Nb and Nb$_{0.44}$Ti$_{0.56}$ are presented in Tab.~\ref{table:1}. Besides the experimental values the equilibrium lattice parameters $a_{\mathrm{eq}}$ are also presented.  For the pure \textit{bcc}-Nb, $a_{\mathrm{eq}}$ is determined by relaxing the crystal structure at a series of different pressures. This is done using projector-augmented plane wave method (PAW)~\cite{blochl} as
implemented in the VASP code~\cite{kresse1996efficiency, kresse1996efficient}. Different combinations of PAW and exchange-correlation (XC) potentials were tried. 
In VASP integration in the Brillouin zone is done on a $12 \times 12 \times 12$ $\Gamma$-centered grid of uniformly distributed $k$-points.  The plane-wave energy cutoff is taken to be \SI{430}{\eV}, and the convergence of the structural optimization is assumed when the total energy change is less than \SI{e-8}{\eV}, and the forces on the atom are less than \SI{e-3}{\eV/\angstrom}.
We find that treating $s$ and $p$ states as semi-core states together with the Perdew–Burke–Ernzerhof (PBE)~\cite{PBE} XC potential yields the structural parameters closest to the experimentally observed ones.

Electronic structure computations are also performed using various standard DFT exchange-correlation functionals such as LDA and GGA.  
The convergence in EMTO is checked on various $k$-mesh sizes up to $69 \times 69 \times 69$ $k$-points in the irreducible part of the first Brillouin zone, although the saturation of the results (no significant change in the density of states, $N(E_F)$) can be noticed for smaller number of $k$-points and the broadening as small as $\delta \approx 10^{-3}$.
To study pressure effects using multiple scattering type electronic structure techniques, such as EMTO, details of the Green's function computations are essential. We use an extended $spdf$ basis set and a broad contour of \SI{3.5}{\rydberg}, including $4s$, $4p$ states. At the same time  60 energy-points on the contour were chosen. 

\begin{table}[h!]
    \begin{center}
        \begin{tabular}{ c|c|c|c}
         & \multicolumn{2}{c}{Nb} & \multicolumn{1
         }{|c}{Nb$_{0.44}$Ti$_{0.56}$} \\
         \hline
         $P$ (\si{\giga\pascal}) & $a_{\mathrm{exp}}$ (\si{\angstrom})  & $a_{\mathrm{eq}}$ (\si{\angstrom}) & $a_{\mathrm{exp}} (\si{\angstrom})$ \\
         \hline
         0 & 3.3089 & 3.3045 & 3.2643 \\
         50 &3.0932 & 3.0936 & 3.0111 \\
        100 &2.9666 & 2.9665 & 2.8747 \\
        150 &2.8801 & 2.8804 & 2.7767 \\
        200 &2.8168 & 2.8155 & 2.7000 \\
        250 &2.7642 & 2.7622 & 2.6412 \\
         \hline
        \end{tabular}
        \caption{Experimental values of the lattice parameters are taken from ~\cite{gu.li.19}, the relaxed lattice parameters are obtained using the DFT-GGA (PBE) exchange correlation functional.}
        \label{table:1}
    \end{center}
\end{table}

\section{Results}

Before discussing electronic correlations and pressure effects in the Nb$_{0.44}$Ti$_{0.56}$ alloy we revisit the electronic structure of pure Nb. We analyze the pressure dependence of the electronic densities of states and the corresponding Fermi surfaces, which allows certain analogies with the Fermi surfaces in the presence of Ti.  
Local dynamic correlations are considered only for Ti (Sec.~\ref{sec:nb-ti}) since the $4d$ Nb states are less correlated in comparison with the $3d$ states of Ti. We estimate the effective mass enhancement as function of the strength of the Coulomb parameter $U$, temperature $T$, and pressure $P$. Finally, we compare the Fermi surfaces of Nb and the Nb$_{0.44}$Ti$_{0.56}$ alloy (Sec.~\ref{sec:FS-results}). 

\begin{figure}[h]
\includegraphics[width=\linewidth,clip=true]{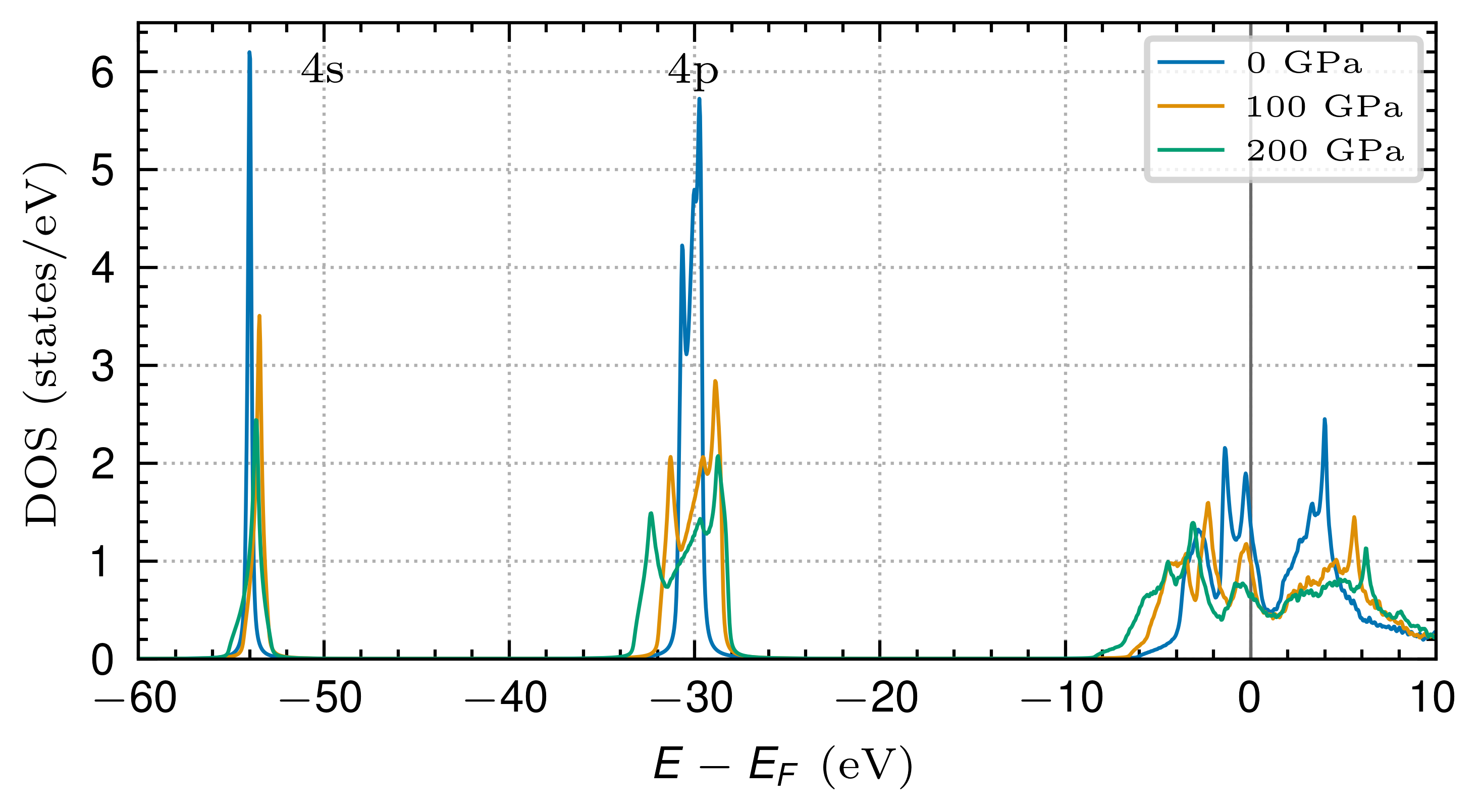}
\includegraphics[width=\linewidth,clip=true]{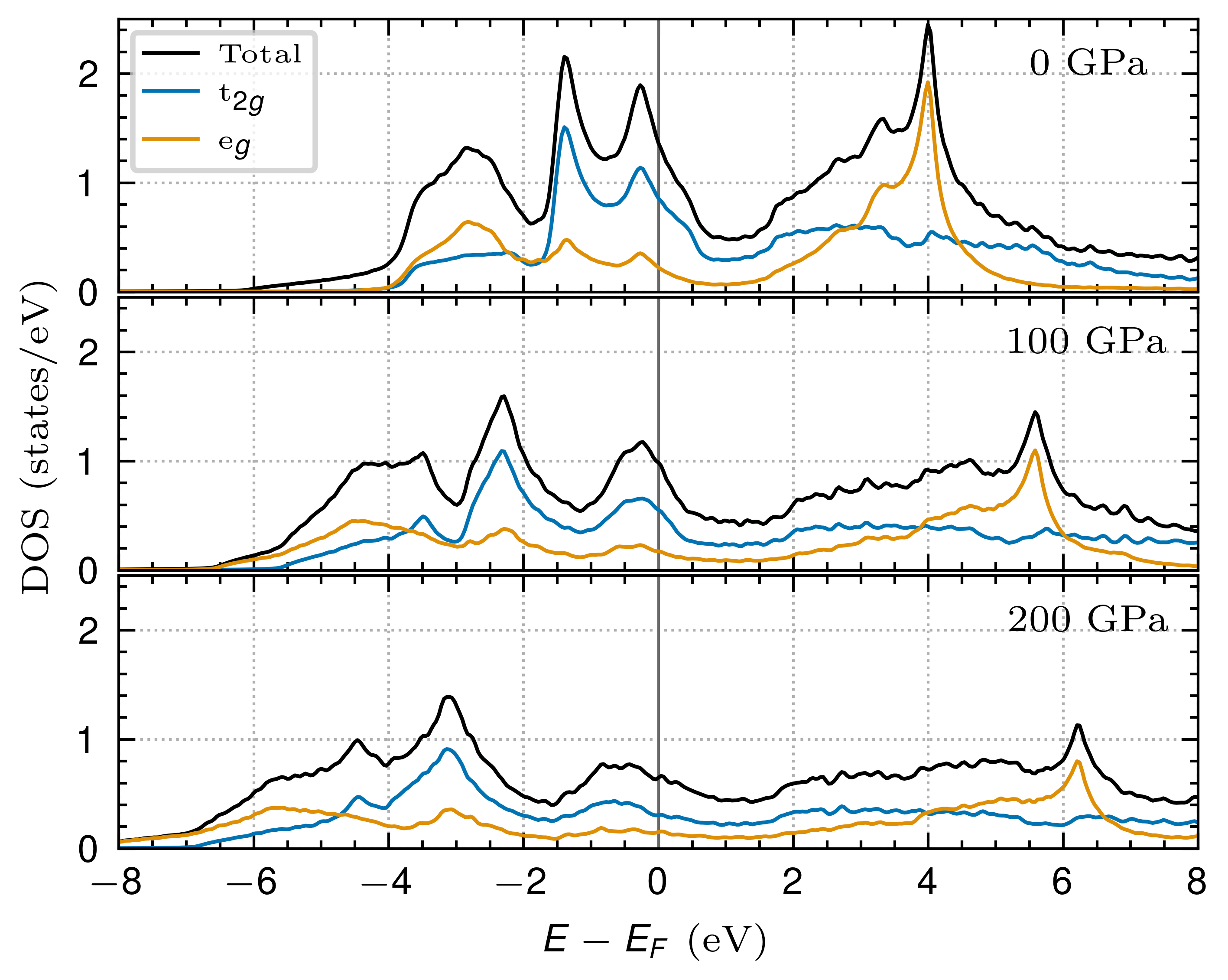}
\caption{Total electronic density of states of pure Nb computed using LDA including the core 4s and semi-core 4p states.
Lower panel: Orbital-resolved $4d$ (t$_{2g}$, e$_g$) electronic density of states of pure Nb. Note the decrease in magnitude with increasing pressure.
}\label{fig:nb_dos_el}
\end{figure}

The electronic structure of niobium was analyzed starting from the early seventies~\cite{matt.70,bu.sm.77}.
According to these studies pure Nb contains five conduction electrons (Nb electronic configuration $4d^4 5s^1$) distributed within three bands: the first is filled, the second almost filled, and the third more than half filled. 
At high pressure, not only the conduction band acquires a larger bandwidth, but also the semi-core states ($4p$) start to change their shape. This is illustrated in the upper panel of Fig.~\ref{fig:nb_dos_el}.
Note that the presence of semi-core states stabilizes the total energy computations, despite the fact that their positions are located at higher binding energies ($\simeq -\SI{30}{\eV}$) with respect to the Fermi level.

It is expected that the most important physical properties are determined by the electronic states in the vicinity of the Fermi level. 
Early experimental studies of photoemission from niobium reported three well-developed peaks at \SI{0.4}{\eV}, \SI{1.1}{\eV} and \SI{2.3}{\eV} below $E_F$~\cite{east.69}. These peaks are fairly well described by our electronic structure calculations. 
The lower panel of Fig.~\ref{fig:nb_dos_el} shows the orbital-projected $4d$ densities of states (t$_{2g}$, e$_g$) in an energy window $E_F \pm \SI{8}{\eV}$. 
Within the cubic \textit{bcc}-structure the t$_{2g}$ and e$_g$ orbitals are orthogonal, and therefore no hybridization between them takes place.
The total DOS around and below $E_F$ is dominated by the (t$_{2g}$, e$_g$) contributions, the ratio of these slightly decreases with increasing pressure. At the same time, a continuous suppression of the density of states in the vicinity of the Fermi level is visible. Our results show that all orbitals respond similarly to pressure, and no signature of orbital selectivity was notable. A comparative analysis of the Fermi surfaces of pure Nb is presented in Sec.~\ref{sec:FS-results}.

\subsection{Pressure and electronic correlations effects in the Nb$_{0.44}$Ti$_{0.56}$ alloy}

\label{sec:nb-ti}
Due to the numerical challenges associated with DMFT computations, in the following we utilize a reduced basis set comprising of $spd$ states, and we limit the contour for calculating the EMTO-Green’s function to \SI{1}{\rydberg}. To ensure the reliability of our results, we compared these calculations with those obtained using the parameters
as detailed in Sec.~\ref{sec:comp_details}. The comparison confirmed that this simplification does not significantly affect the density of states (DOS) within the energy range of $E_F \pm \SI{10}{\eV}$.

\begin{figure}[h!]
\includegraphics[width=\linewidth,clip=true]{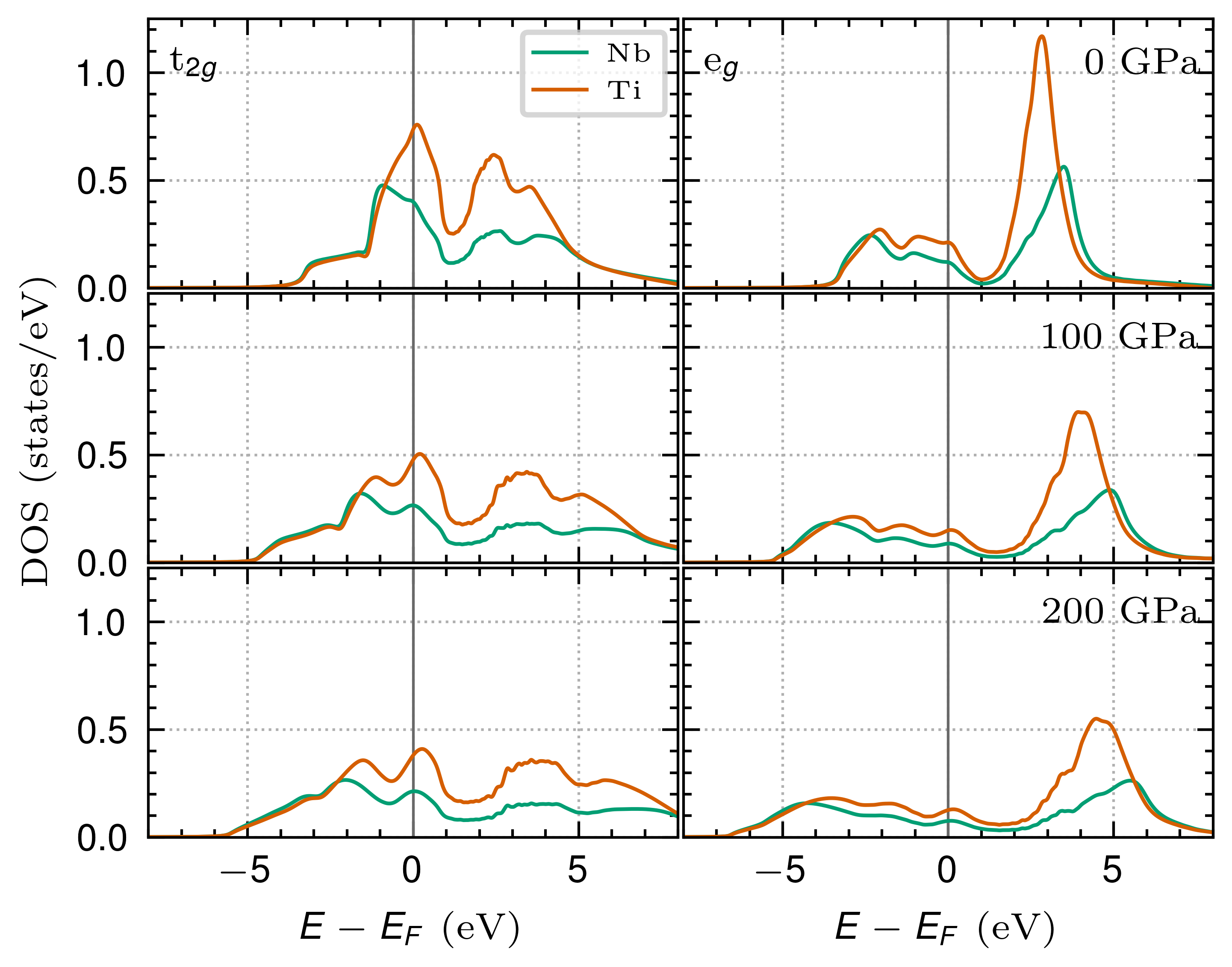}
\caption{Orbital-resolved electronic densities of states of Nb$_{0.44}$Ti$_{0.56}$ at different pressures, for the non-interacting case,  $U = J = 0$.}
\label{fig:nbti_dos_orbs}
\end{figure}

To start with, Fig.~\ref{fig:nbti_dos_orbs} presents the orbital-resolved DOS of Nb$_{0.44}$Ti$_{0.56}$ up to \SI{200}{\giga\pascal} for the non-interacting case ($U = 0$, $J = 0$). 
Due to the cubic symmetry of the system, no mixture (hybridization) occurs between the t$_{2g}$ and e$_g$ orbitals, resulting in Green's functions that are diagonal in orbital space. The Ti t$_{2g}$ orbitals dominate the DOS at $E_F$ for all pressures, while the Nb t$_{2g}$ orbitals also contribute significantly in the energy range $-\SI{2}{\eV} \le E_F \le \SI{1}{\eV}$. Note that the contribution of e$_g$ orbitals near $E_F$ is consistently at least four times smaller than that of t$_{2g}$ orbitals at any pressure.
Thus, our computations strengthen the previous conjecture that electronic states contributing to superconductivity should have a dominant t$_{2g}$ character associated to both Nb-$4d$ and Ti-$3d$ electrons. 

\begin{figure}[h!]
\includegraphics[width=\linewidth,clip=true]{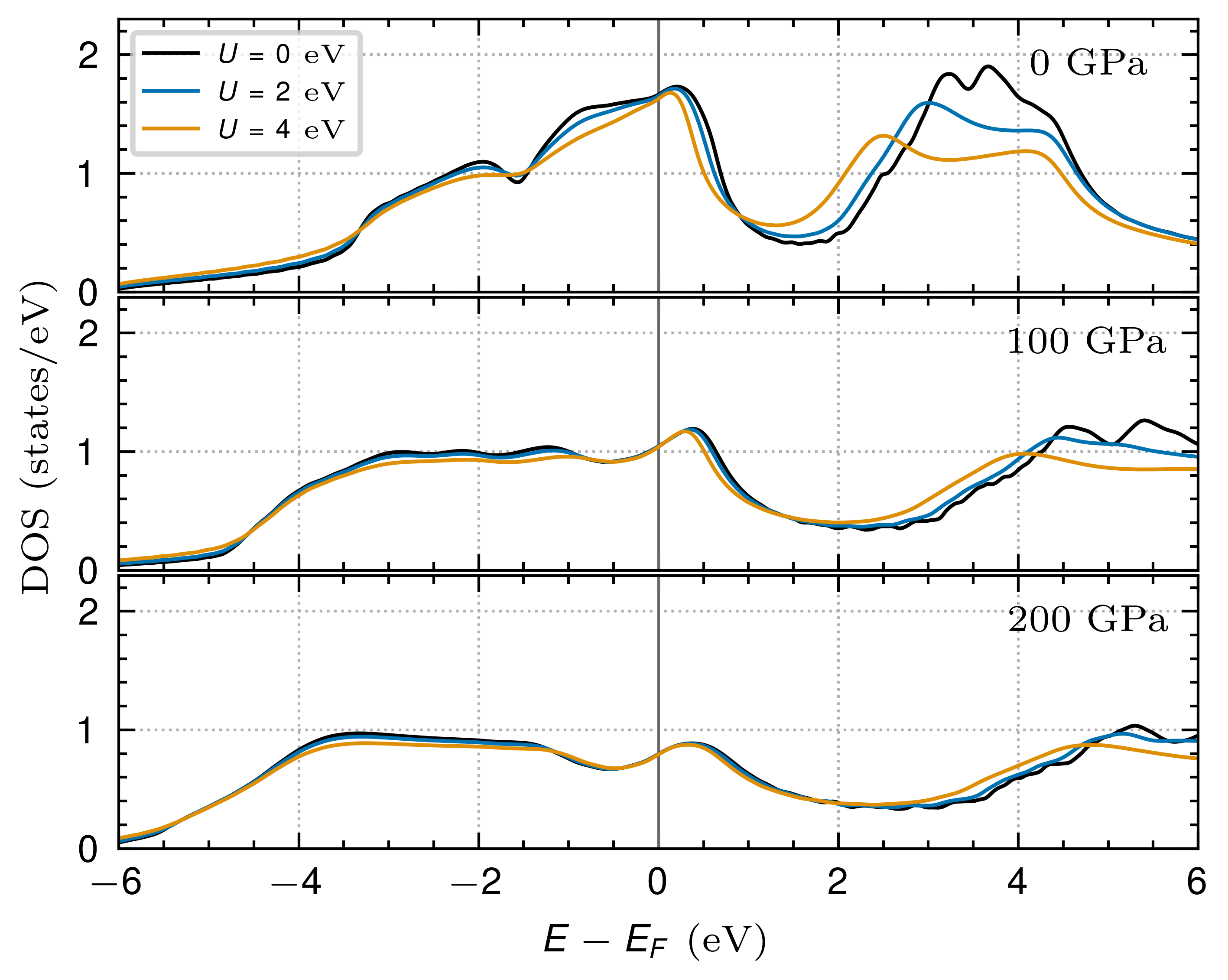}
\caption{Total electronic densities of states of Nb$_{0.44}$Ti$_{0.56}$ computed using LDA+DMFT at different pressures and interaction strengths $U,J$ as described above.}\label{fig:nbti_dos_dmft}
\end{figure}

At the (non-interacting) LDA level, the spectral features of Nb$_{0.44}$Ti$_{0.56}$ alloys exhibit a broader profile compared to pure Nb, a consequence of the imaginary component of the CPA self-energy. 
Figure~\ref{fig:nbti_dos_dmft} shows the results for the total DOS with and without including the DMFT corrections at different pressures.
For calculations with $U$ values larger than \SI{2}{\eV}, we employ a constant $J = \SI{0.6}{\eV}$, while for $U < \SI{2}{\eV}$, we maintain a fixed ratio of $U/J = 2/0.6 \approx 3.33$ to ensure the positivity of the effective Coulomb repulsion $U - 3J$. Increasing the Hund’s coupling up to $J = \SI{0.9}{\eV}$ did not produce any significant changes in the spectral functions.

Including correlation effects, additional broadening appears due to the imaginary part of the DMFT self-energies. 
Compared to the non-interacting case, electronic correlations induce a gradual suppression of the DOS near the Fermi level and promote the emergence of tails at the band edges. For a range of $U$ parameters up to \SI{5}{\eV}, the value of DOS($E_F$) remains pinned, highlighting the Fermi-liquid nature of the electronic states.
Overall, the spectral functions exhibit minimal changes in the presence of electronic correlations (see Fig.~\ref{fig:nbti_dos_dmft}), 
nevertheless, some degree of anisotropy in the superconducting order parameter is anticipated as a result of the different symmetry of the orbitals within the crystal structure.

\begin{figure}[h!]
\includegraphics[width=\linewidth,clip=true]{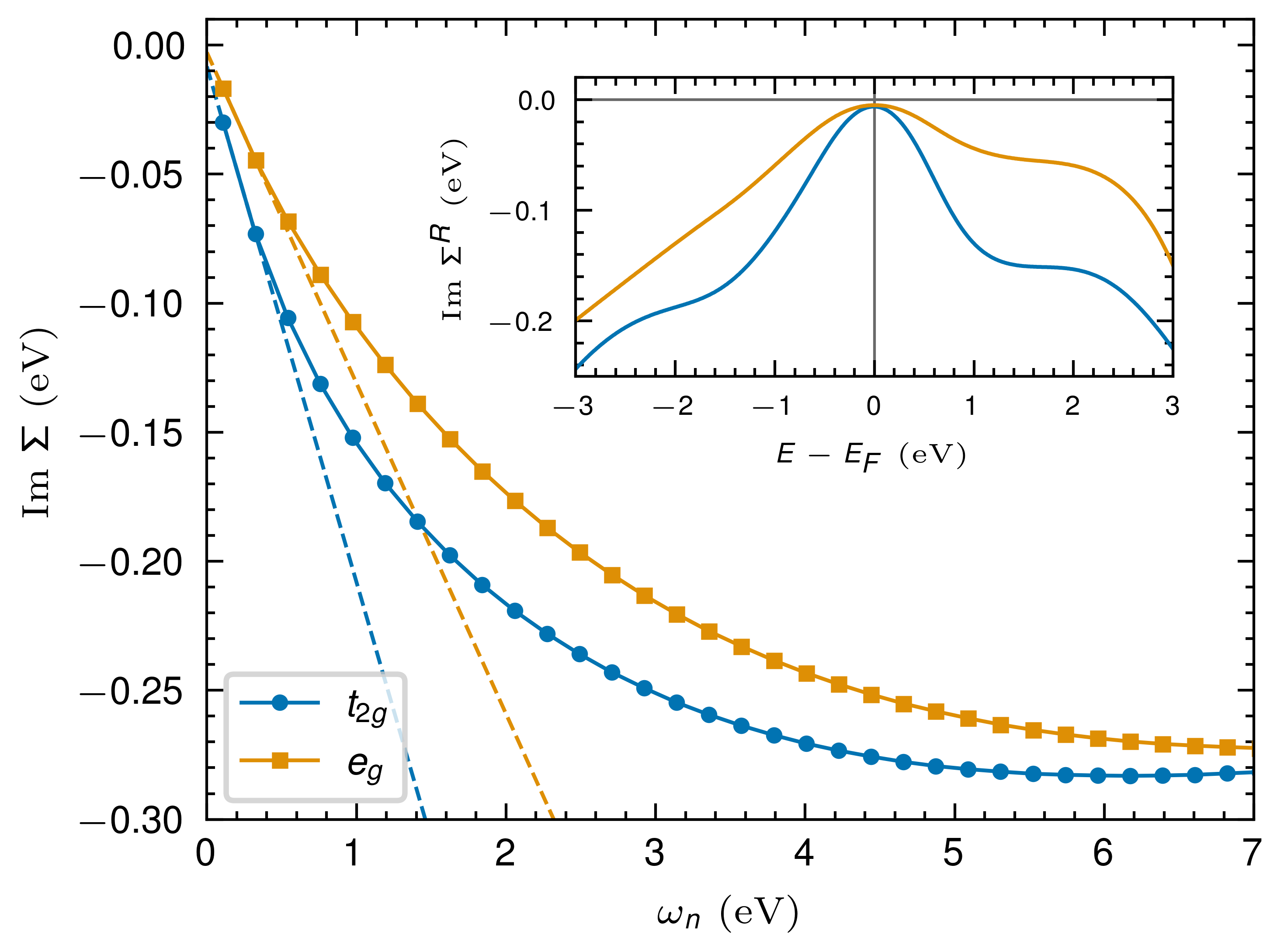}
\caption{Imaginary part of the self-energies as a function of the Matsubara frequencies, $\omega_n$, of the Ti t$_{2g}$ and e$_g$ orbitals in Nb$_{0.44}$Ti$_{0.56}$ as obtained from LDA+DMFT, with $U=\SI{2}{\eV}$ and $J=\SI{0.6}{\eV}$, at \SI{400}{\kelvin} and ambient pressure (\SI{0}{\giga\pascal}). The inset shows the parabolic dependence of the imaginary part of the retarded self-energy on energy $E$ around $E_F$, confirming the Fermi-liquid behavior.}\label{fig:Sig_nbti}
\end{figure}

To further characterize the effects of electronic correlations, Fig.~\ref{fig:Sig_nbti} displays the imaginary part of the orbital-resolved self-energies.
The inset shows the analytically continued self-energies, i.e., on the real energy axis, which exhibit a quadratic dependence near the Fermi energy, $\Im \Sigma(E) \propto -(E-E_F)^2$.
Both the t$_{2g}$ and e$_g$ orbitals show Fermi-liquid behavior; however, as expected, the effective masses are different. The analytically continued self-energies are obtained using Pad\'e-approximants methods~\cite{vi.se.77,we.ot.20}.
 
However, in order to avoid possible inaccuracies of the analytical continuation, we focus the discussion on the computational results obtained for the self-energy on the Matsubara axis.
The Fermi-liquid state is characterized by a linear dependence of the imaginary part of the self-energy:
\begin{align}
    \Im\Sigma_{(l,m)}^\sigma(\omega_n) \simeq - \big[(Z_{(l,m)}^\sigma)^{-1} - 1\big] \omega_n    
\end{align} 
where $Z_{(l,m)}^\sigma$ denotes the quasiparticle spectral weight of electrons in orbital $(l,m)$ with spin $\sigma$.
At low Matsubara frequencies, $\Im \Sigma_{\mathrm{t}_{2g}/\mathrm{e}_g}(\omega_n)$ approaches zero linearly (dashed lines in Fig.~\ref{fig:Sig_nbti}) for both orbitals.
This behavior indicates that the t$_{2g}$ and e$_g$ electrons act as long-lived quasiparticles, with their scattering rate vanishing in the vicinity of the Fermi surface---a signature for the existence of a Fermi-liquid metallic state.
Across the entire temperature range of \SI{100}{\kelvin} to \SI{400}{\kelvin}, the linearity of the imaginary part of the self-energy is preserved. These findings provide robust evidence for the Fermi-liquid nature of the electronic system.

Within DMFT, the quasiparticle mass enhancement $m^{\star}/m$, relative to the band mass $m$, becomes local and can be determined in the zero-temperature limit as $m^{\star}/m = 1 - \mathrm{Im} \Sigma(\omega_n) / \omega_n|_{\omega_n\to 0}$. This quantity is commonly computed numerically in quantum Monte Carlo (QMC) simulations by evaluating the slope of the imaginary part of the self-energy at a few low Matsubara frequencies~\cite{ge.ko.96}. The dashed lines in Fig.~\ref{fig:Sig_nbti} represent the slopes for the t$_{2g}$ and e$_g$ orbitals.

\begin{figure}[h!]
\includegraphics[width=\linewidth,clip=true]{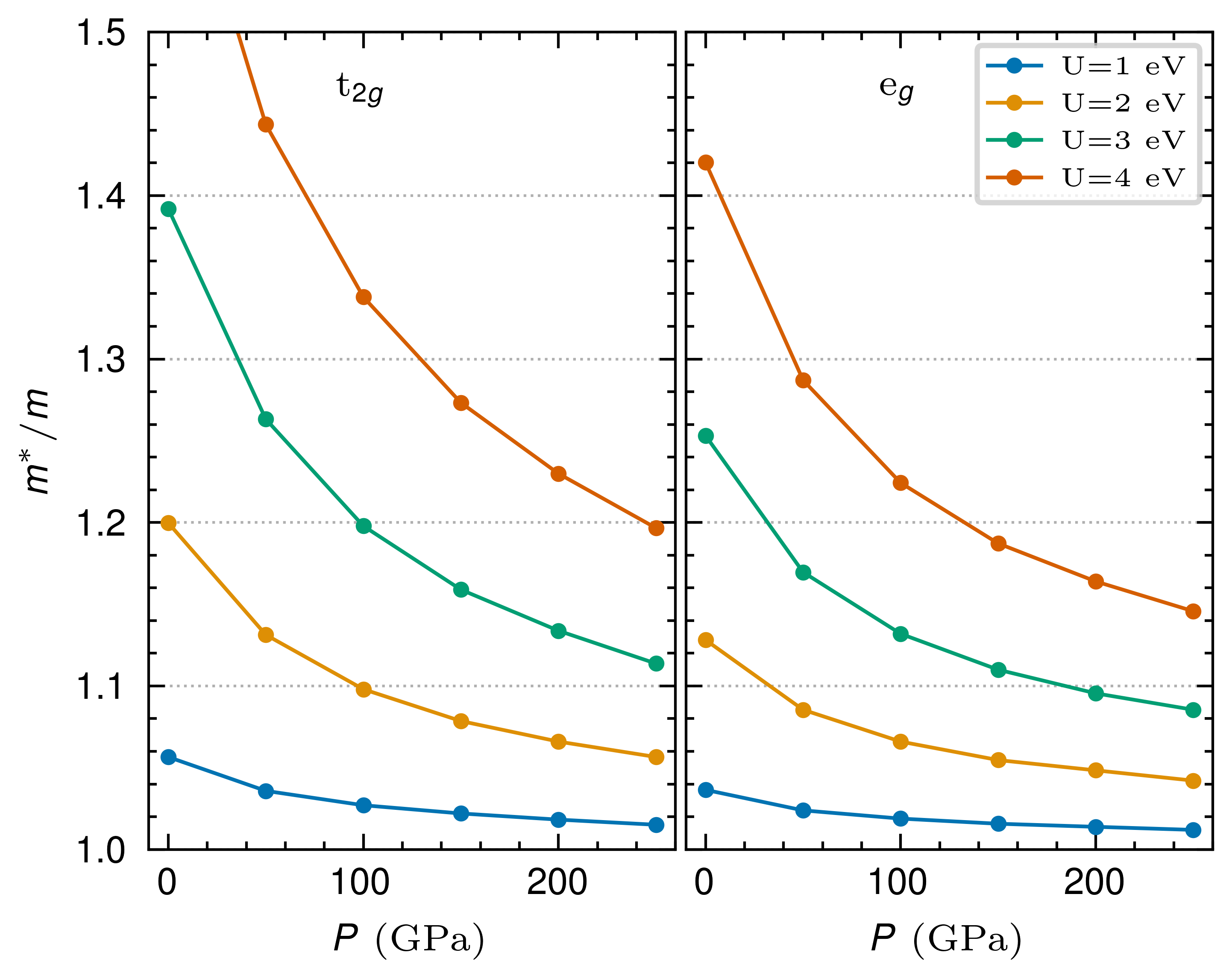}
\includegraphics[width=\linewidth,clip=true]{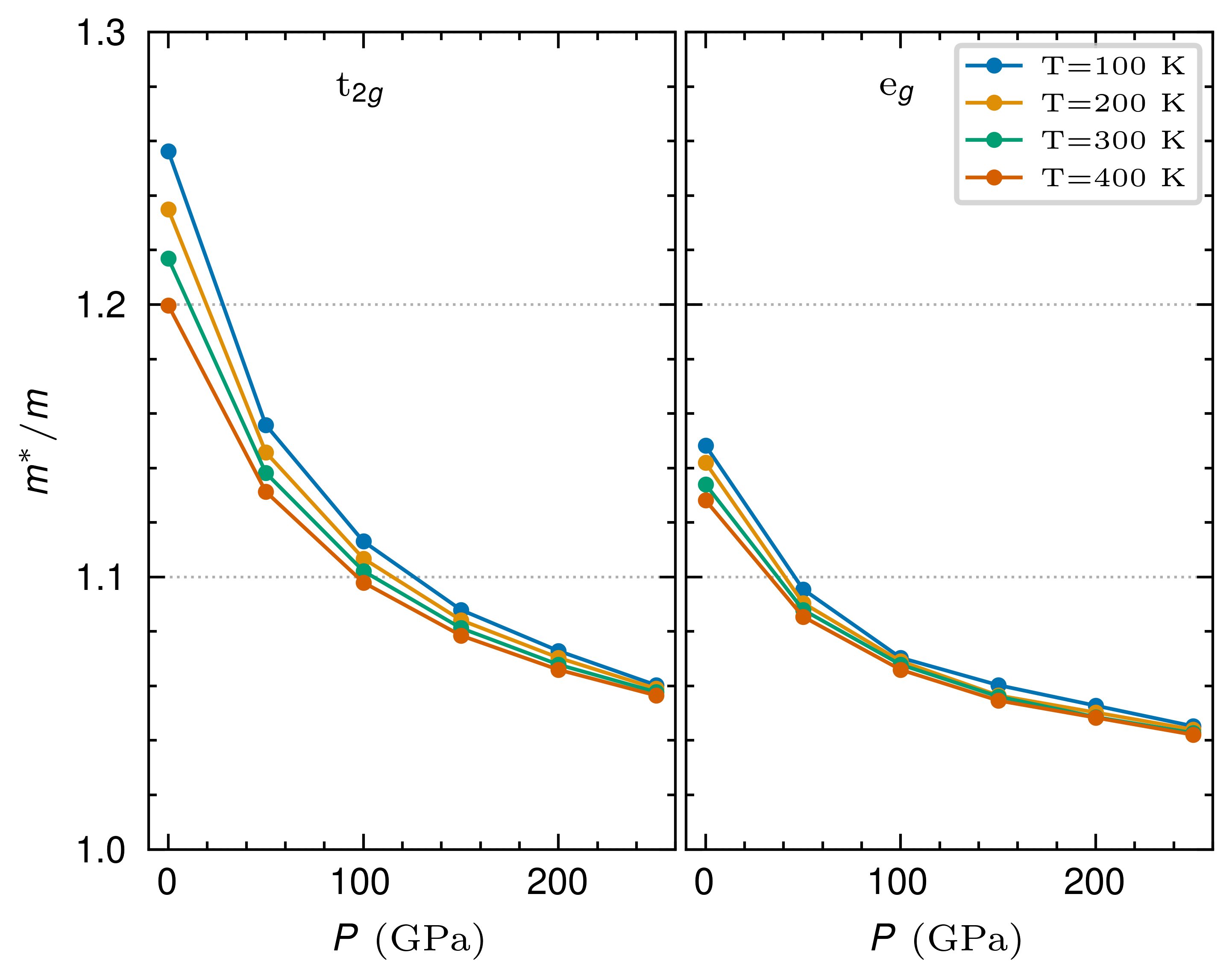}
\caption{Upper/lower panel: Pressure dependence of the effective mass renormalization of the t$_{2g}$ and e$_g$ orbitals for various $U/T$ values. For the choice of $J$, see text.}
\label{fig:meff_nbti}
\end{figure}

The corresponding $m^{\star}/m$ are shown in Fig.~\ref{fig:meff_nbti} for various Hubbard parameters, up to $U = \SI{5}{\eV}$.
At all temperatures, the effective mass of both orbitals increases with increasing $U$. While the effective mass at higher temperatures is somewhat reduced, this overall trend is maintained, thereby confirming the conclusion that correlation effects are diminished with increasing pressure. 
We observe no evidence suggesting disparate/separate/distinct reactions of the orbitals to pressure or electronic correlations. Consequently at extremely high pressures, it is reasonable to surmise the absence of significant anisotropic effects in the superconducting order parameter, and the anomalous Green's function for t$_{2g}$ and e$_g$ electrons should exhibit a similar behavior. Inter-orbital superconducting pairing is not to be expected.

The computed effective mass at various temperatures is presented in the lower panel of Fig.~\ref{fig:meff_nbti}. As pressure increases, a continuous suppression of $m^{\star}/m$ is observed. In particular, at \SI{250}{\giga\pascal}, the effective mass approaches its band value.
The variation in the effective mass with pressure is a consequence of the broadening of the electronic spectral function due to the energy spread of the electronic states. This tendency is consistent across all interaction strength parameters and temperatures.

\subsection{Fermi surfaces of Nb and Nb-Ti alloys}
\label{sec:FS-results}
\begin{figure*}
\includegraphics[width=0.32\linewidth,clip=true]{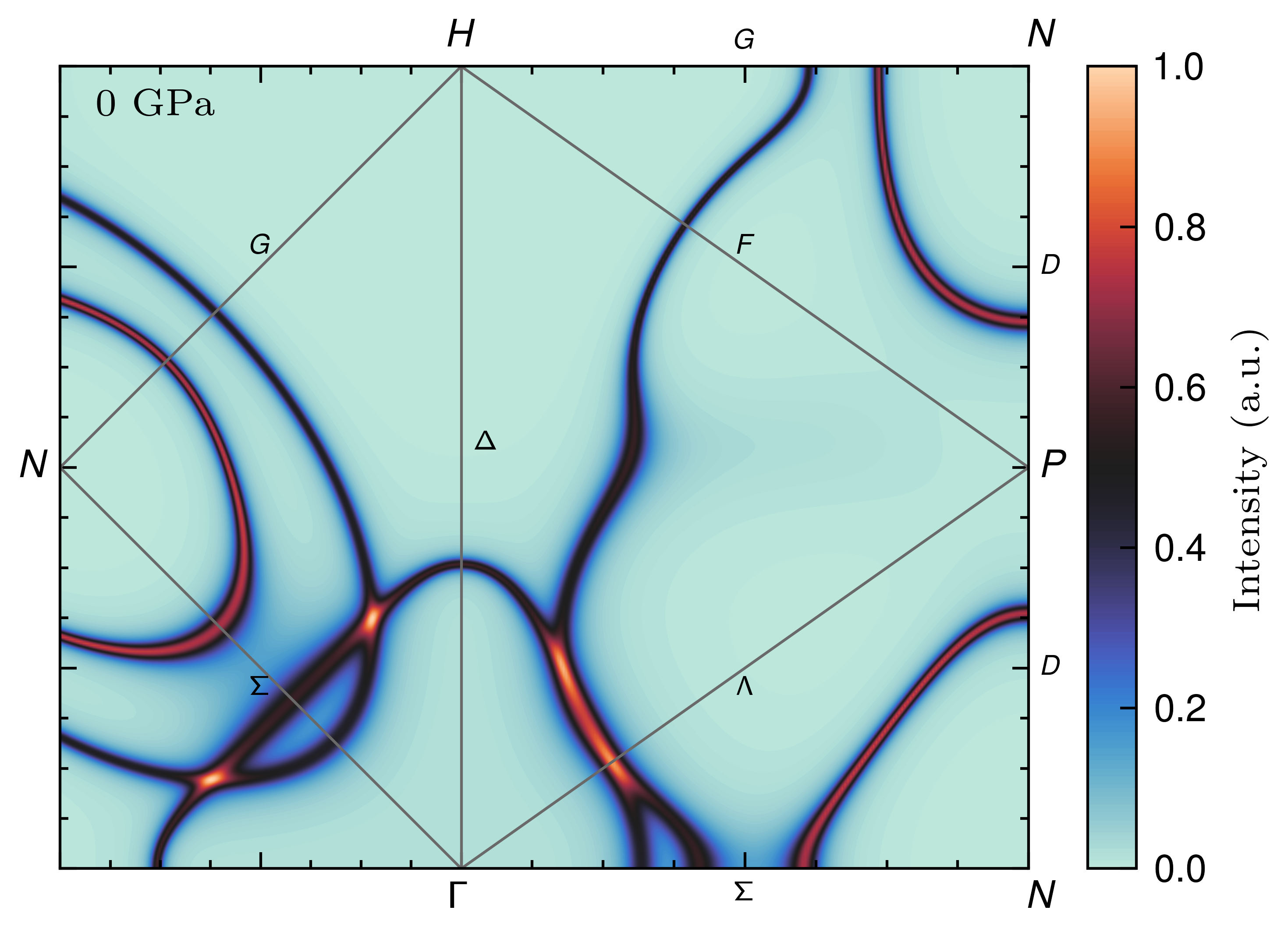}
\includegraphics[width=0.32\linewidth,clip=true]{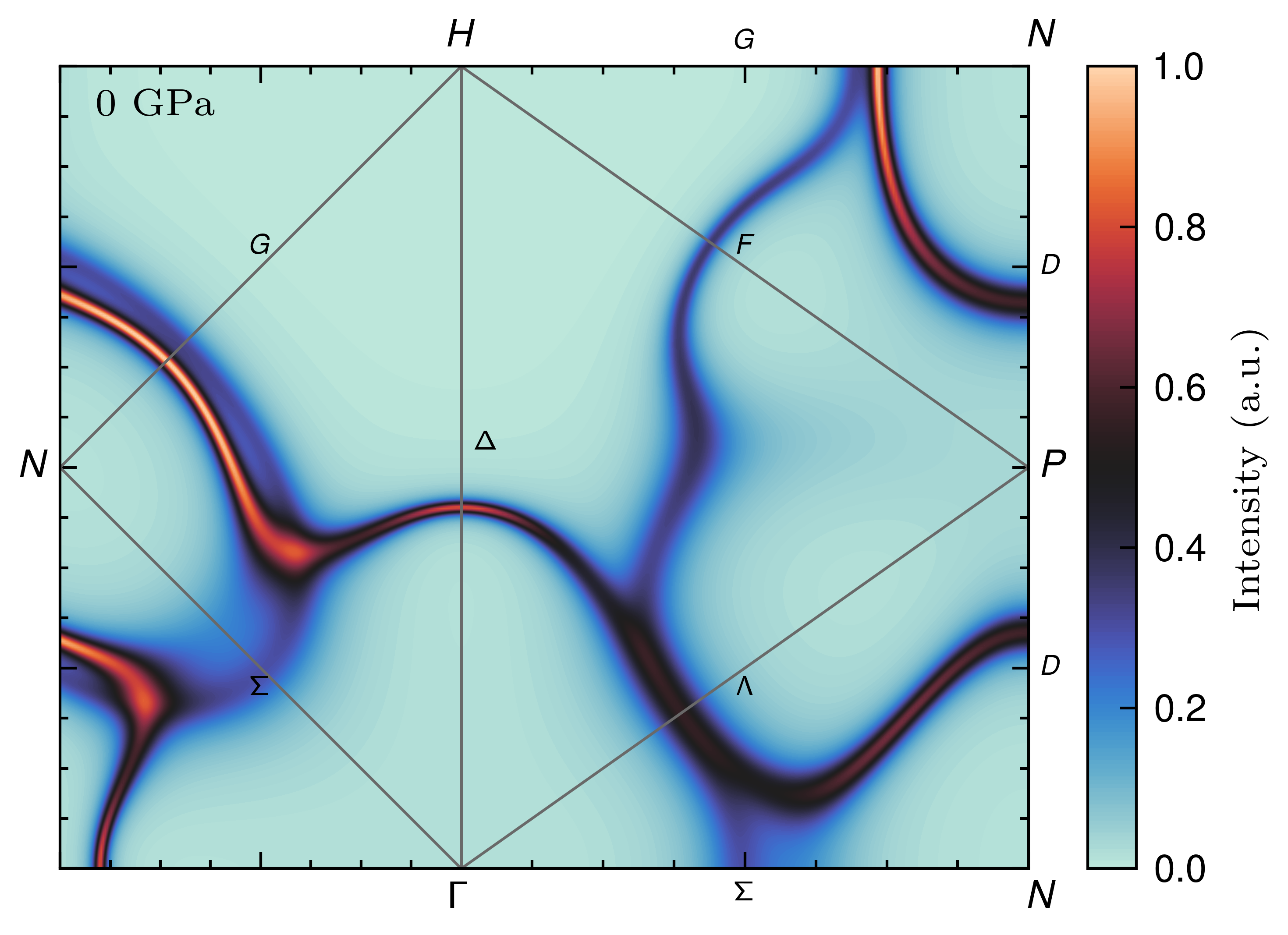}
\includegraphics[width=0.32\linewidth,clip=true]{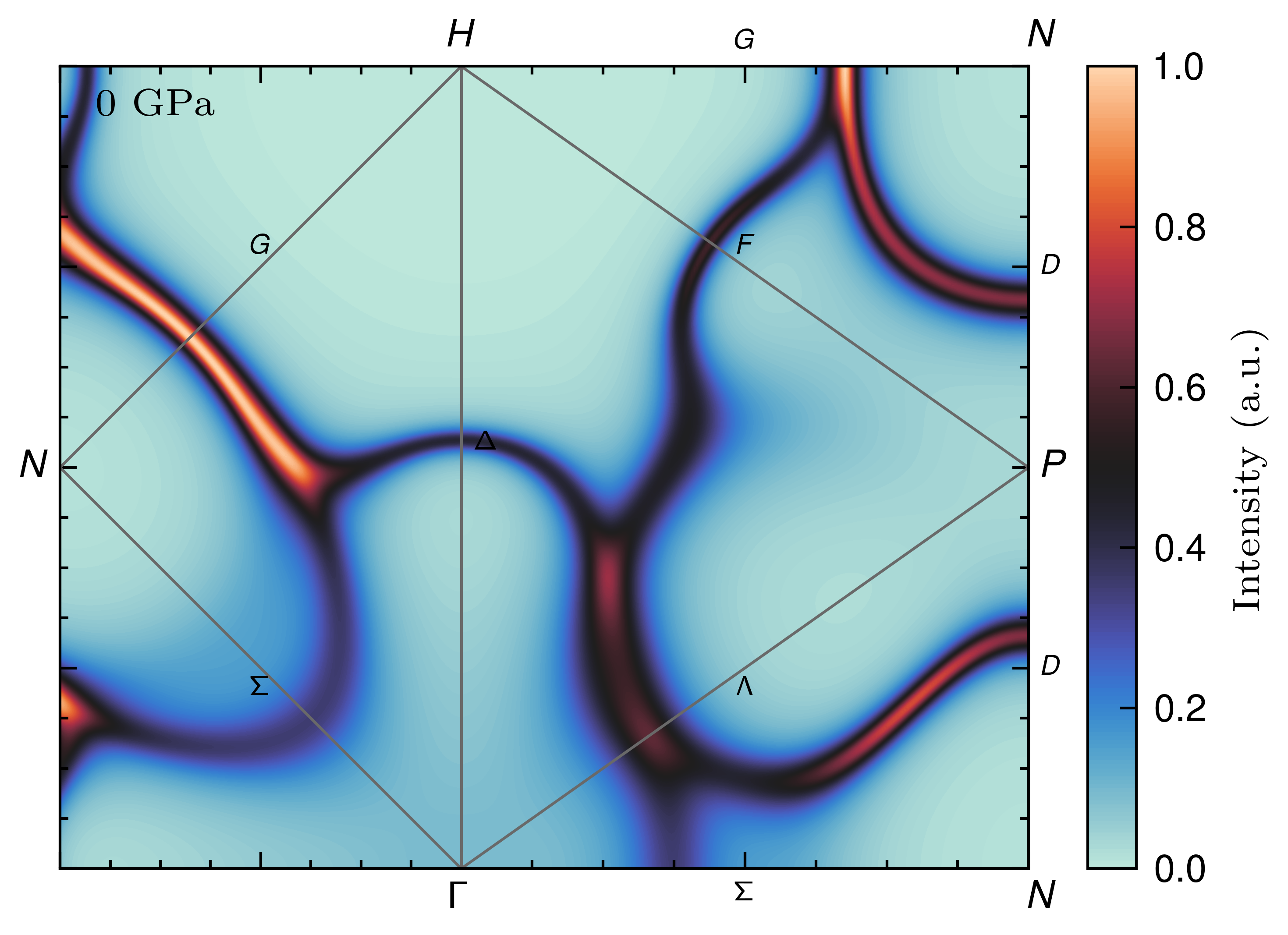}\\
\includegraphics[width=0.32\linewidth,clip=true]{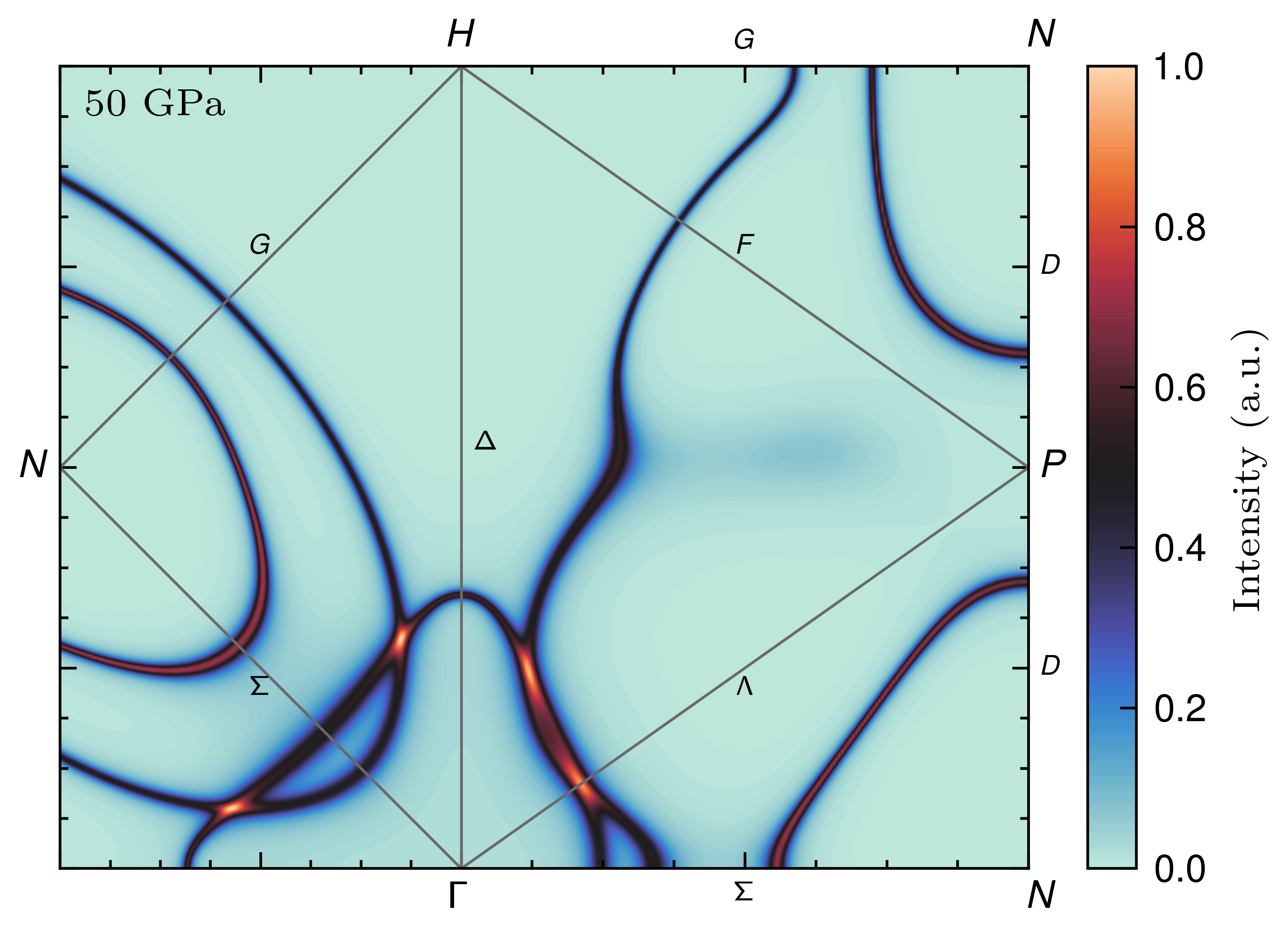}
\includegraphics[width=0.32\linewidth,clip=true]{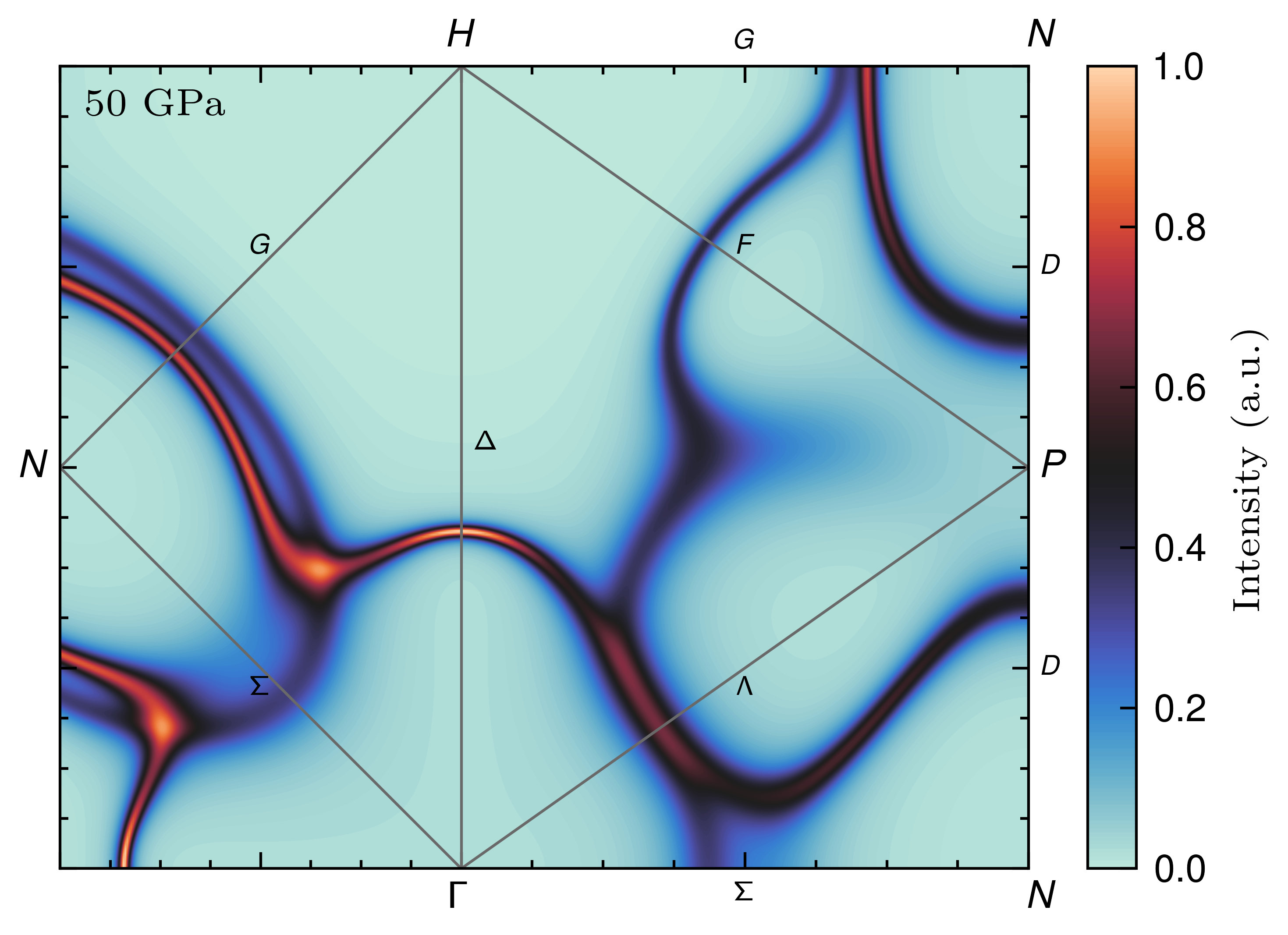}
\includegraphics[width=0.32\linewidth,clip=true]{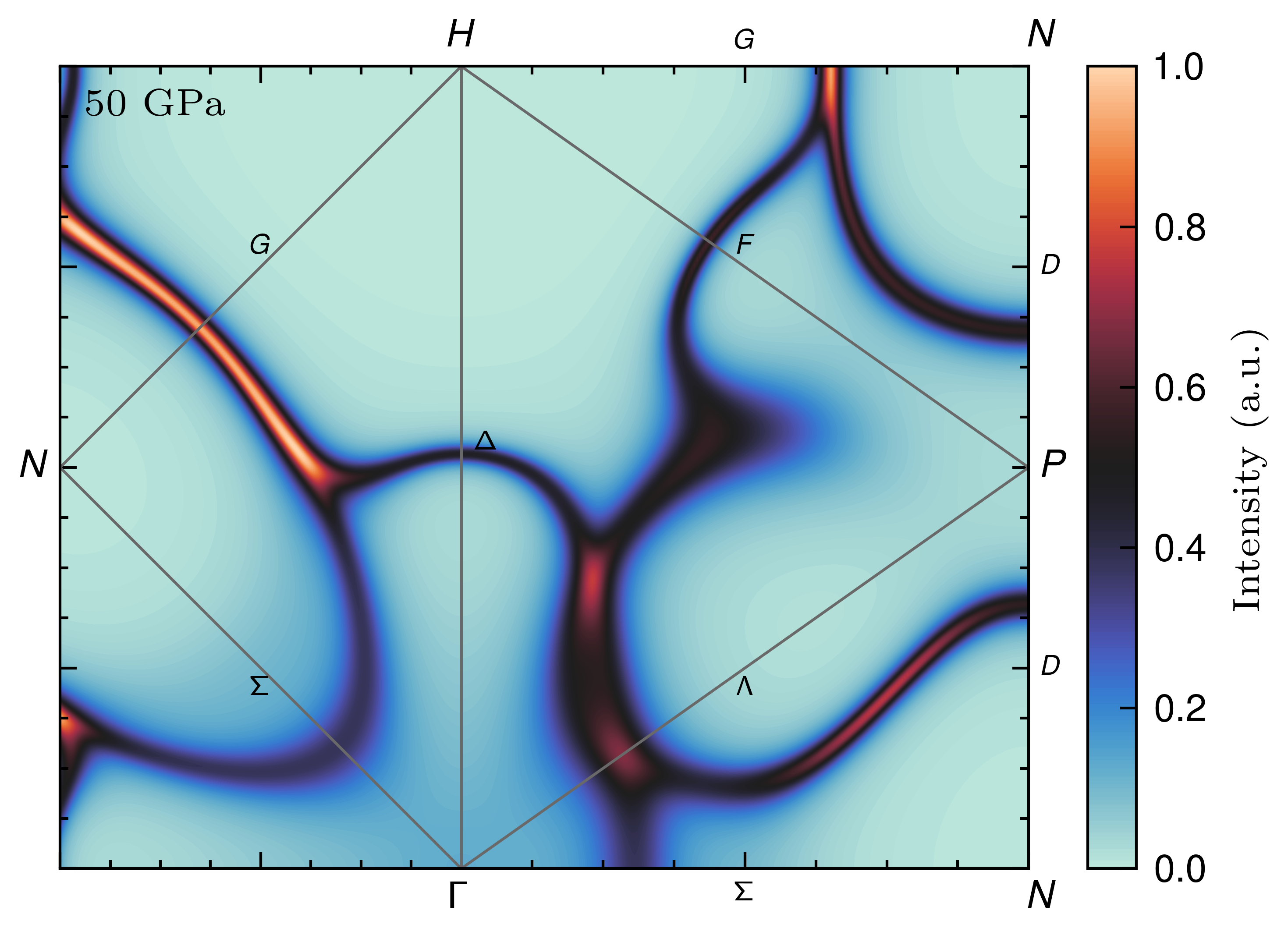}\\
\includegraphics[width=0.32\linewidth,clip=true]{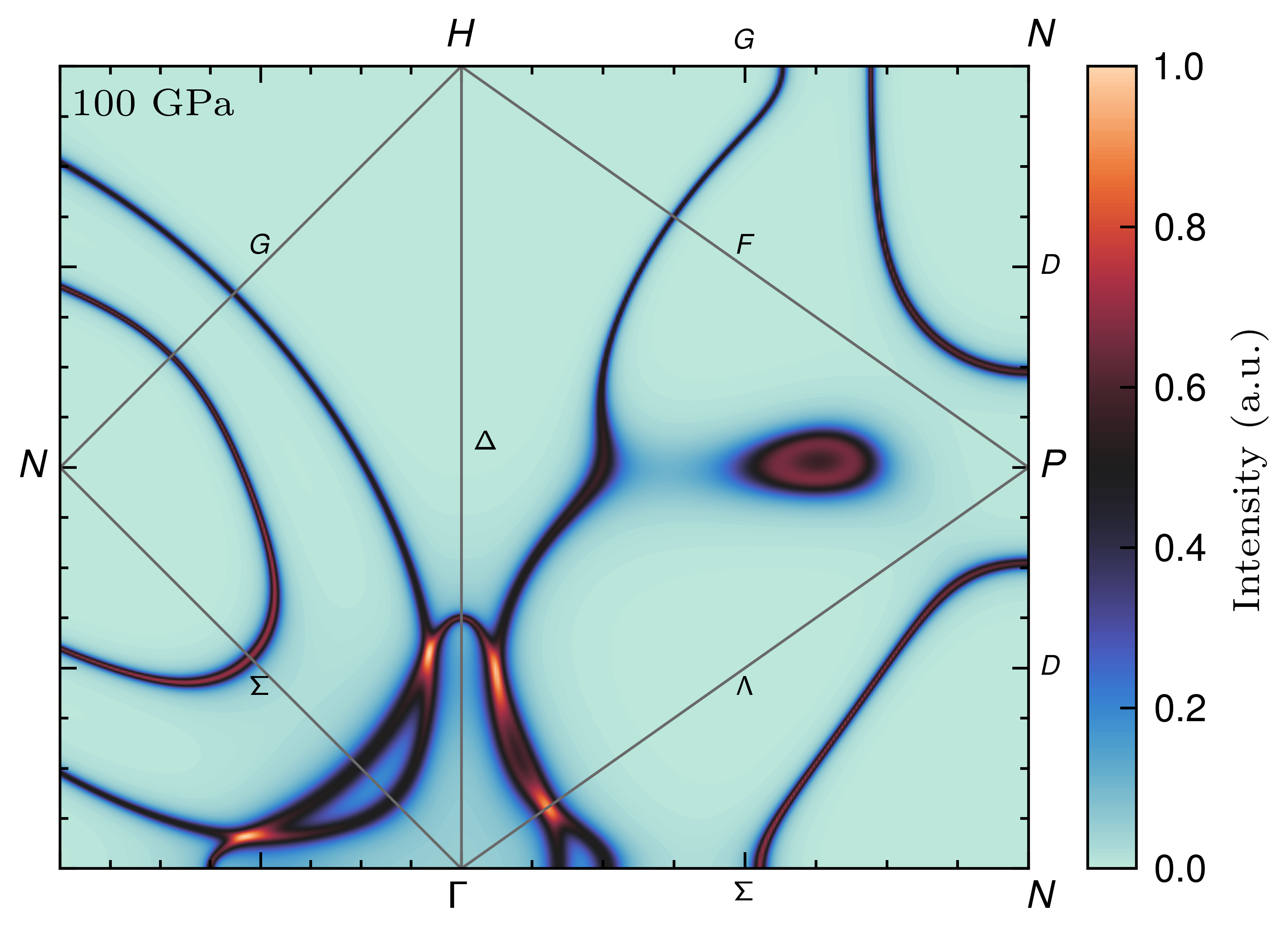}
\includegraphics[width=0.32\linewidth,clip=true]{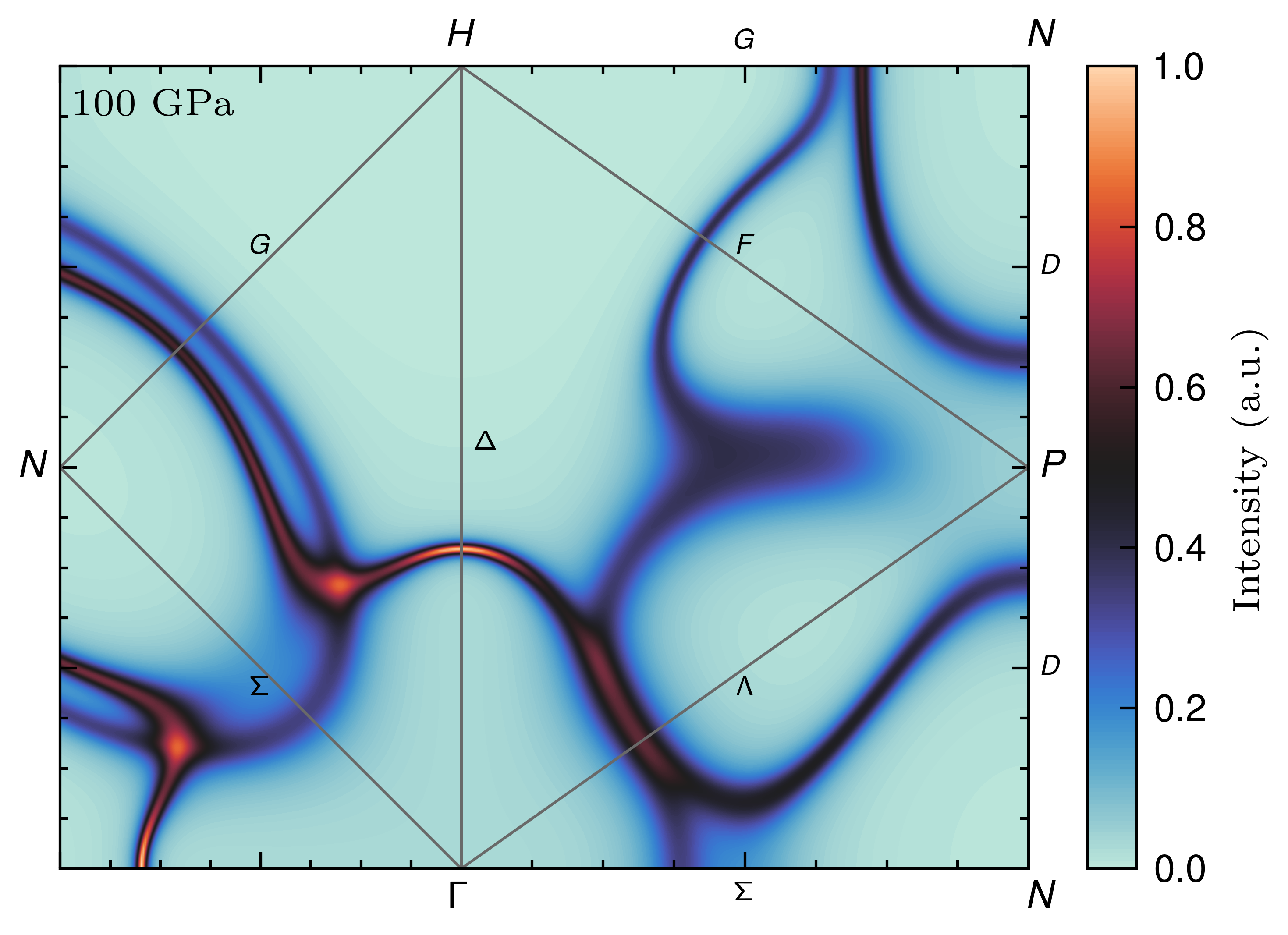}
\includegraphics[width=0.32\linewidth,clip=true]{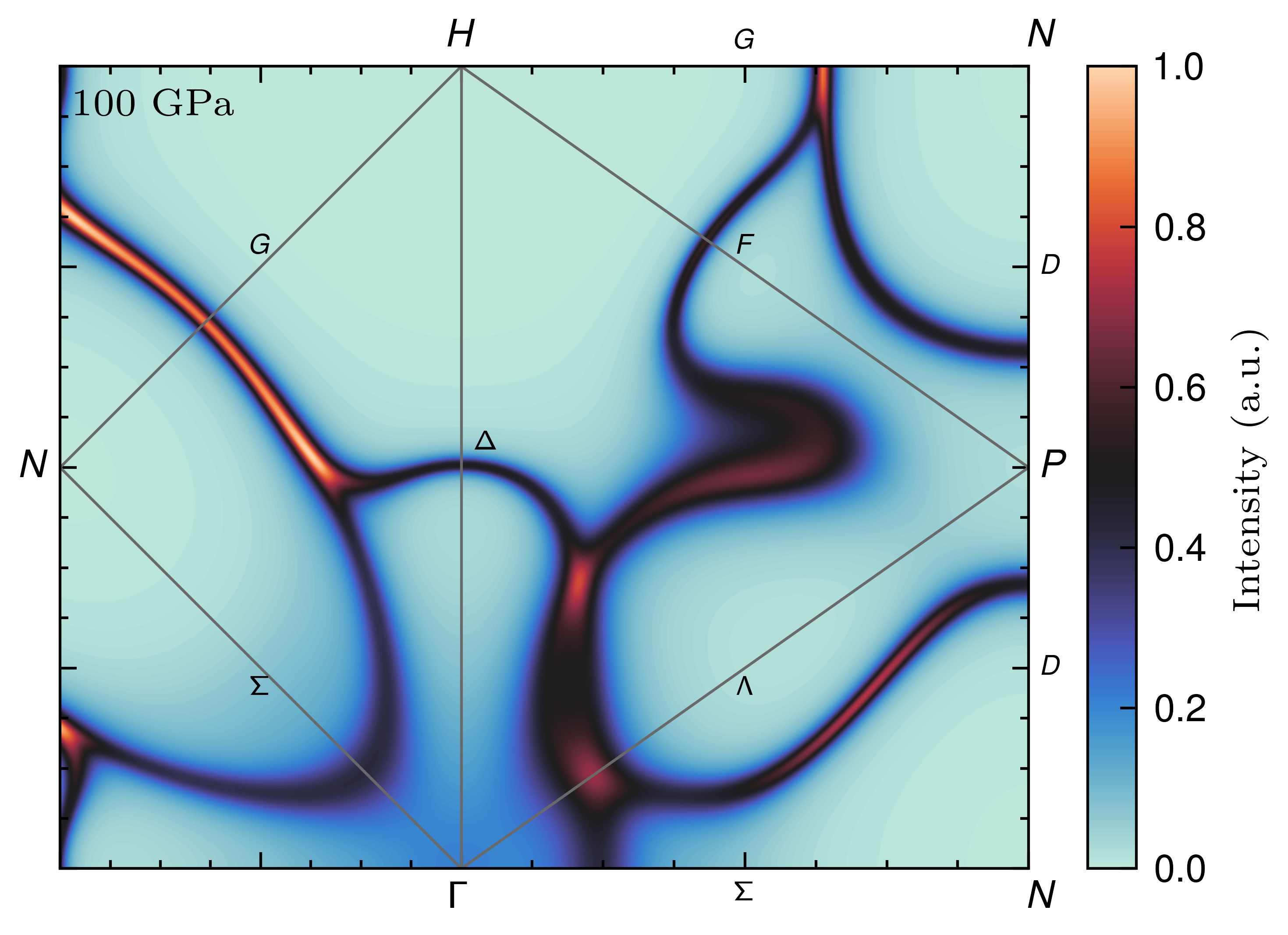}\\
\includegraphics[width=0.32\linewidth,clip=true]{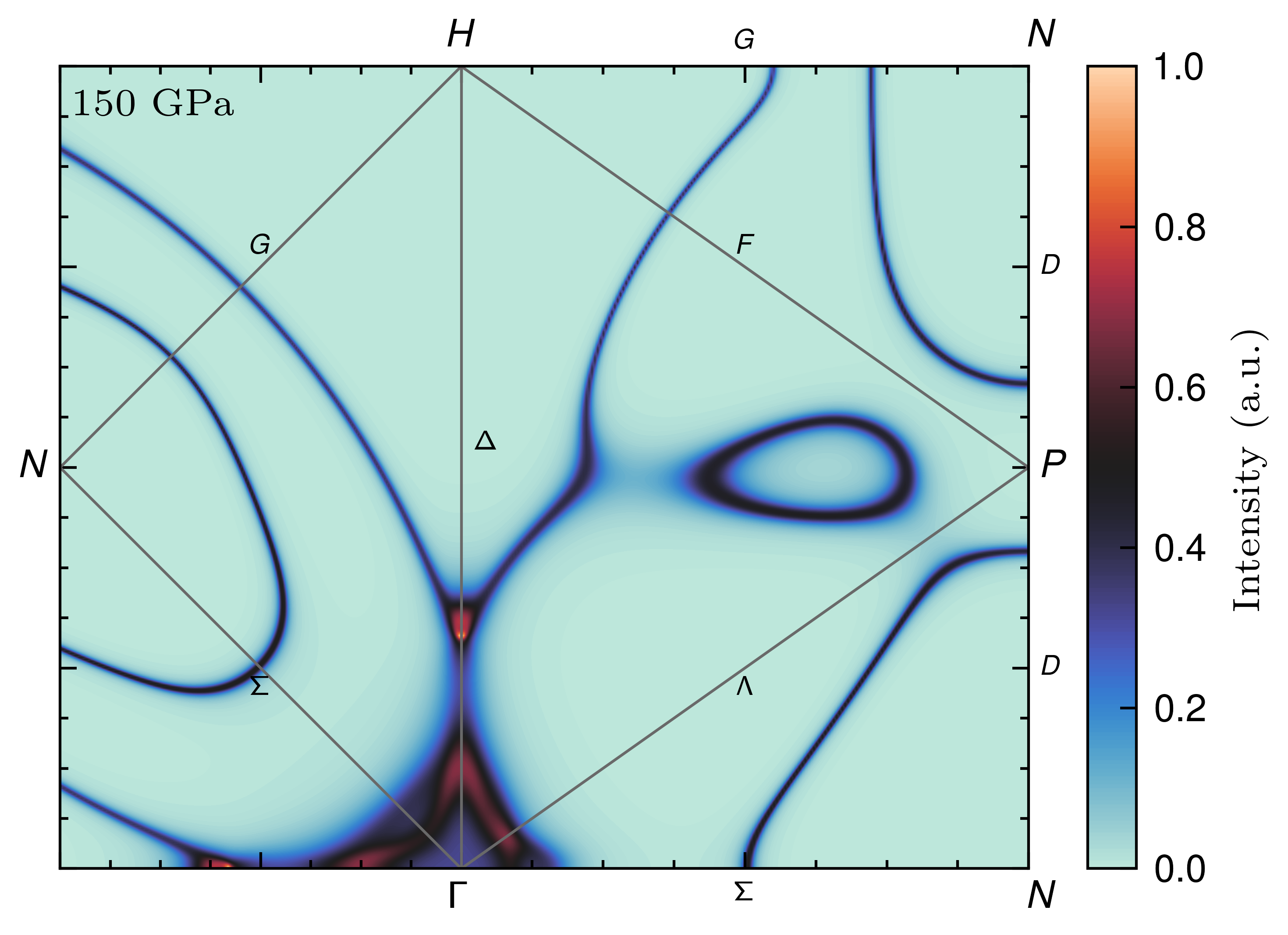}
\includegraphics[width=0.32\linewidth,clip=true]{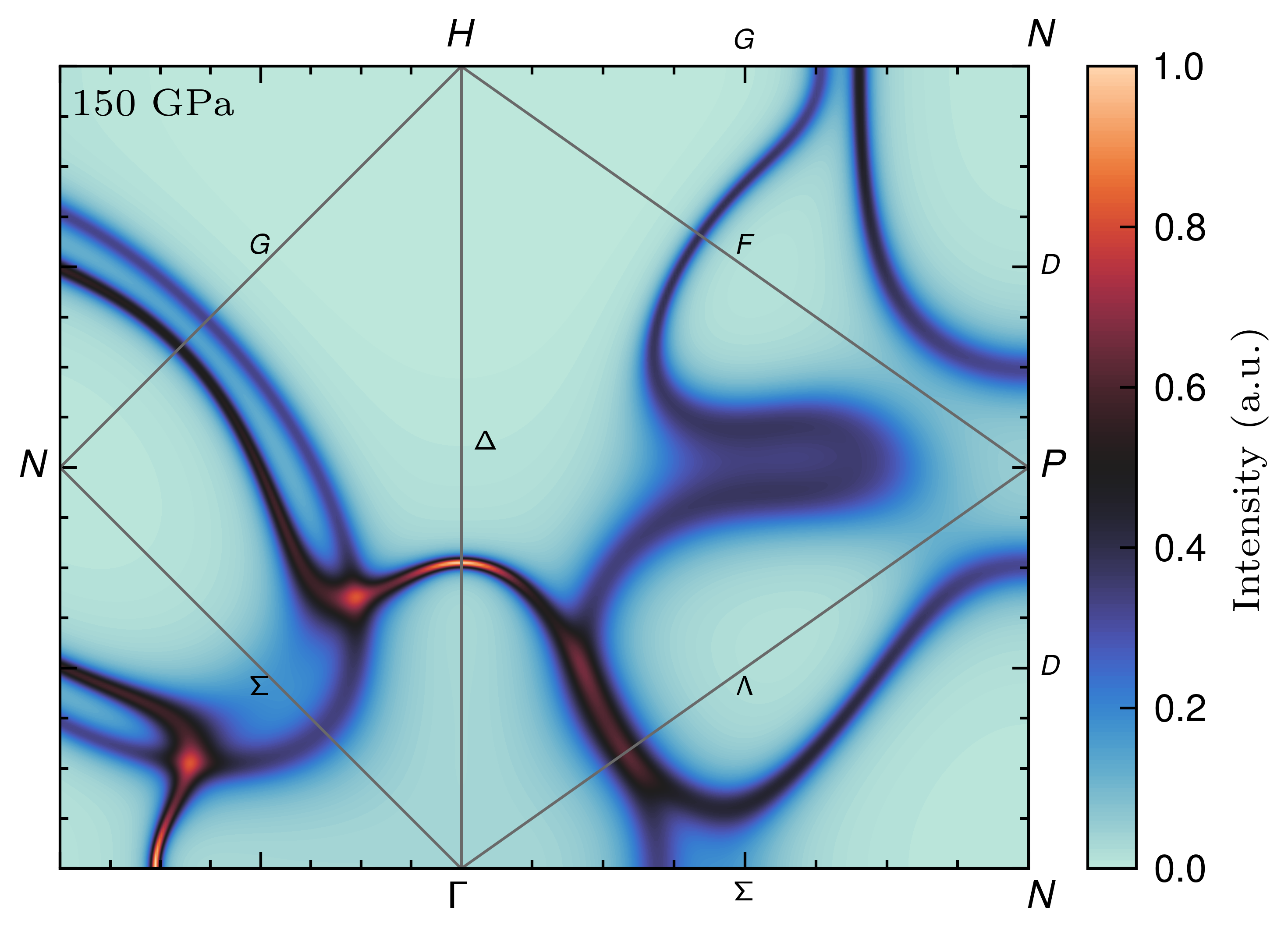}
\includegraphics[width=0.32\linewidth,clip=true]{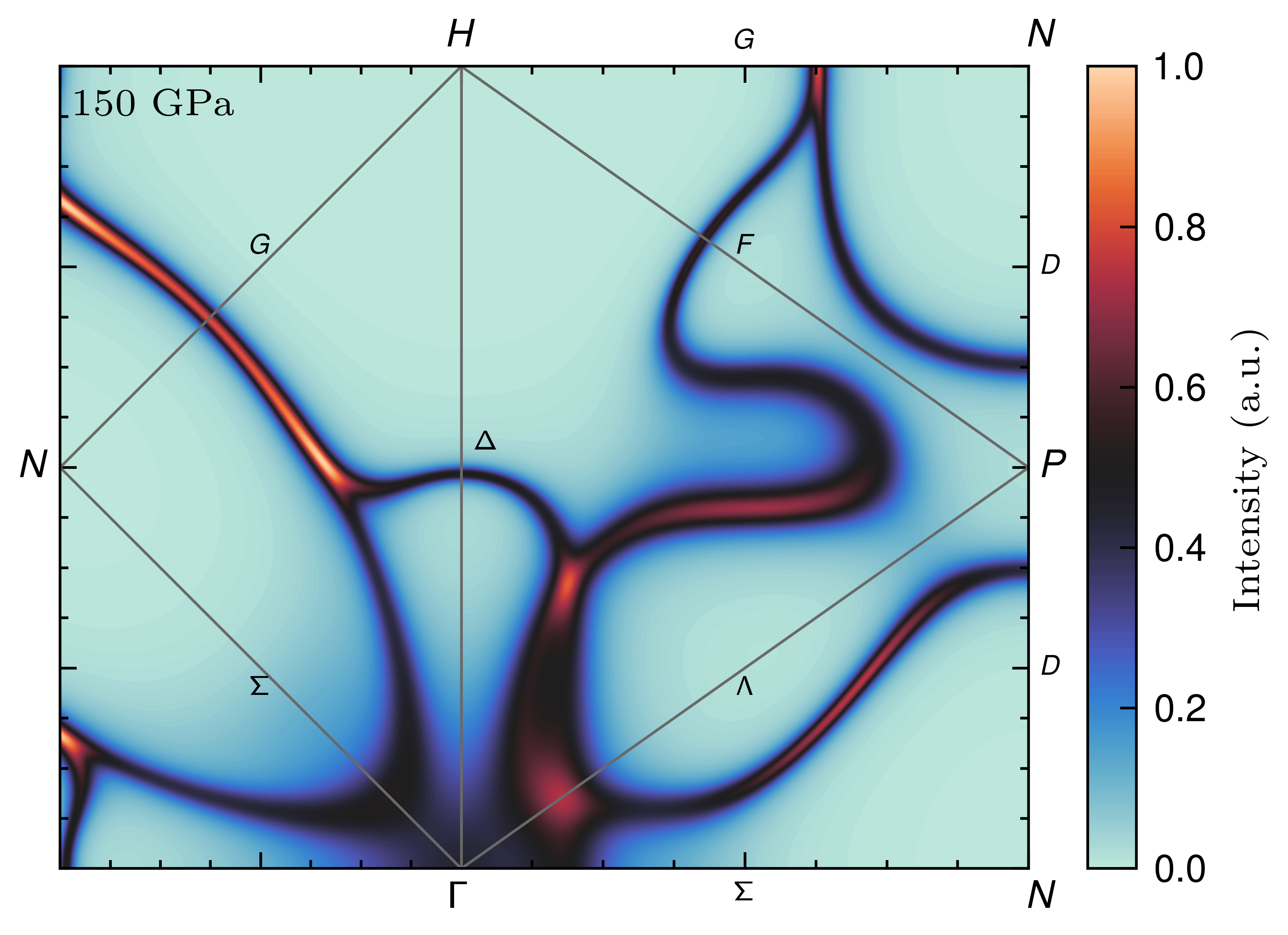}
\caption{
    The evolution of the Fermi surface with pressure. 
    The first column represents the LDA Fermi surface of bulk Nb at ambient pressure, \SI{50}{\giga\pascal}, \SI{100}{\giga\pascal} and \SI{150}{\giga\pascal}. The second and the third column, respectively, show the Fermi surfaces for the Nb$_{0.44}$Ti$_{0.56}$ alloy using LDA(+CPA) and LDA(+CPA)+DMFT ($U=\SI{2}{\eV}$, $J=\SI{0.6}{\eV}$, $T=\SI{400}{\kelvin}$).
    The symmetry points in the first Brillouin zone of the body centered cubic (\textit{bcc}) lattice are given by 
    $\Gamma=(0,\ 0,\ 0)$, $H=(0,\ 0,\ 2\pi/a)$, $P=(\pi/a, \ \pi/a,\ \pi/a)$, and $N=(0,\ \pi/a, \ \pi/a)$, where $a$ is the lattice constant.
}\label{fig:FS_Nb-Ti}
\end{figure*}

Figure~\ref{fig:FS_Nb-Ti} shows the central cross-sections of the Fermi surface of pure Nb and Nb-Ti at various pressures.
The triangular section $\Gamma HN$ corresponds to the $(100)$ plane, while the rectangular region $\Gamma HNPN$ represents the central $(110)$ plane.

The Fermi surfaces of pure Nb are shown in the left column, with increasing pressure from top to bottom. A significant pressure effect upon the shape of the Fermi surface is apparent: 
At ambient conditions (zero pressure), a hole-like FS is observed at the zone center (the $\Gamma$ point), exhibiting a distorted octahedral shape commonly referred to as the ``jack''. At the $N$ points, a set of two distorted ellipsoids is observed, along with additional hole sheets extending from $\Gamma$ to $H$ along the $\langle 100 \rangle$ direction, known as the ``jungle-gym''. The octahedra and the jungle-gym (the external ellipsoid centered at $N$) touch twice within the triangular $(100)$ plane, with high-intensity points along a direction perpendicular to $\Sigma$. These contact points are symmetrically positioned with respect to the $\Gamma N$ direction, and are located at approximately $(0, \ -0.22, \ 0.62) \ \pi/a$ and $(0, \ -0.62, \ 0.22) \ \pi/a$.
Another contact point occurs along the $\Gamma$-$P$ direction (also referred to as the $\Lambda$-direction), which extends within the plane as a high-intensity line.

Calculations of the pressure effect on Nb at \SI{6}{\giga\pascal} and \SI{26}{\giga\pascal} do not predict any significant changes in the Fermi surface topology~\cite{an.pa.73,an.pa.81,st.ti.97}.
Based on augmented plane wave calculations~\cite{el.ko.77}, it was shown that the $N$-$H$ branch of the third band crosses the Fermi level only under pressure, potentially accounting for the disappearance of the neck between the ellipsoids and the jungle-gym. Overall, the reduction of $T_c$ observed above \SI{70}{\giga\pascal} was attributed to the Fermi level entering a region of lower density of states~\cite{bo.pa.77,st.ti.97}. 
According to our results, increasing the pressure leads to a continuous reduction in the size of the octahedron-shaped feature around the $\Gamma$-point. 
Within the $(100)$ plane, the two ellipsoidal structures grow in size, and their contact points gradually separate. At \SI{150}{\giga\pascal}, the size of the octahedra is significantly reduced.
Within the rectangular area $\Gamma HNPN$, a hole band touching the Fermi surface is observed at \SI{50}{\giga\pascal}. This feature evolves into an oval shape at \SI{100}{\giga\pascal}, and its size continues to increase with growing pressure. The center of this newly formed hole band is located at $(0.63 , \ 0.63, \ 1.00) \ \pi/a$.

The second and third columns of Fig.~\ref{fig:FS_Nb-Ti} depict the Fermi surface of Nb$_{0.44}$Ti$_{0.56}$, calculated using LDA(+CPA) (middle column) and LDA(+CPA)+DMFT (right column).
The parameters are $U=\SI{2}{\eV}$, $J=\SI{0.6}{\eV}$, and $T=\SI{400}{\kelvin}$.
In contrast to pure Nb, all spectra displayed in the second and third column are broadened. This broadening arises from two distinct sources: the CPA self-energy (for the second column) and, additionally, the DMFT self-energy (third column), the latter accounting for electronic correlations. 
A reduction in the size of the octahedron centered at the $\Gamma$ point with increasing pressure is also observed in the CPA Fermi surfaces. However, the volume reduction is less significant than in pure Nb.

Another noticeable effect involves the ellipsoids centered at the $N$-point. At zero pressure, the external ellipsoid appears diffuse, while the internal one is intense. With increasing pressure, the ellipsoids separate, and the intensity shifts towards the external one. Despite these changes, the coordinates of the contact points between the octahedron and the ellipsoids remain largely unchanged with increasing pressure. 
Similar to pure Nb, starting at \SI{50}{\giga\pascal}, the shape of the jungle-gym structure within the rectangular area $\Gamma HNPN$ is modified. 
An occupied band is promoted towards the Fermi level and becomes visible at $(0.61 , \ 0.61, \ 1.03) \ \pi/a$. However, this band does not form a closed hole area. Instead, the lobes separate into distinct volumes along the $\Lambda$ and $F$ directions, while its ``neck'' region around the $P$-point is reduced in size.

Finally, we provide a brief description of the DMFT Fermi surfaces computed for the parameters $U = \SI{2}{\eV}$, $J = \SI{0.6}{\eV}$, and $T = \SI{400}{\kelvin}$. Notably, no significant qualitative differences in the Fermi surface are observed for other values of $U$, $J$, and $T$.
At zero pressure, the most pronounced change relative to the CPA Fermi surface is observed for the hole volume centered at the $\Gamma$-point, which deforms and increases in size.
However, with increasing pressure, the volume of this feature decreases.
In the triangular section $\Gamma HN$, the ellipsoids around the $N$-point remain connected, and their size increases with pressure, contributing to the suppression of the octahedron at the $\Gamma$-point. 
Within the $(110)$ plane, a more significant distortion of the jungle-gym structure is observed, with its boundaries exhibiting an increased intensity. The lobes along the $\Lambda$ and $F$ directions are elongated, and the contact point along the $\Gamma P$-direction separates from the extended touching line within the $(110)$ plane. Notably, the intensity of this feature increases at higher pressures.

We now turn to a discussion of the Fermi-liquid characteristics of the Fermi surface of Nb$_{0.44}$Ti$_{0.56}$.
According to the Fermi-liquid picture, the low-energy behavior of electrons can be described in terms of interacting quasiparticles, and their collective excitations. The Landau phenomenological theory assumes a one-to-one correspondence between the quasiparticle excitations of the interacting system and the single-particle excitations of the non-interacting electron system.
This correspondence implies that the volume of the FS of interacting quasiparticles is related to the electron density in the same manner as for non-interacting electrons~\cite{land.57}.
Luttinger and Nozieres~\cite{lutt.60,no.lu.62,lu.no.62} justified the Landau assumption using many-body perturbation theory, establishing the relation between the Fermi surface volume and the electron density, commonly referred to as the Luttinger theorem or Luttinger sum rule.
However, deviations from the Luttinger sum rule have been observed in impurity models~\cite{cu.ni.18}, which lead to the distinction between Fermi and Luttinger surfaces.
In our computations, we find that the Luttinger sum rule is satisfied. Technically, the total electron number is determined from the spectral function derived from the one-electron Green's function:
\begin{align}
    N = -\frac{1}{\pi} \sum_{(l,m),{\sigma},\mathbf{k}} \int_{-\infty}^{E_F} \dd E \,\Im{G^{\sigma}_{(l,m)}(\mathbf{k},E+i\eta)} \Big\vert_{\eta\to 0^+}
\end{align}
The interacting retarded Green's function $G^{\sigma}_{(l,m)}(\mathbf{k},z)$ is evaluated along the complex contour $z=E+i\eta$, and incorporates the DMFT self-energy, $\Sigma_{(l,m)}^{\sigma}(z)$. 
Within Landau's theory, quasiparticles are defined as the poles of this Green's function:
\begin{align}
    G^\sigma_{(l,m)}(\mathbf{k},z) \simeq \frac{Z_{(l,m)}^{\sigma}(\mathbf{k},z)}{z-\tilde{\epsilon}_{(l,m)}^{\sigma}(\mathbf{k},z)  +i\Gamma_{(l,m)}^{\sigma}(\mathbf{k},z)} + \text{incoh.},
\end{align}
where $\tilde{\epsilon}_{(l,m)}^{\sigma}$ represents the quasiparticle energy and $\Gamma_{(l,m)}^{\sigma}(\mathbf{k},z)$ the scattering rate (broadening). The incoherent contribution is typically neglected. Within the DMFT approximation, the self-energy is local, resulting in a quasiparticle weight $Z$ and a relaxation rate $\Gamma$ that are independent of momentum.  
As illustrated in Fig.~\ref{fig:Sig_nbti}, the conditions for Fermi-liquid behavior are satisfied around $E_F$. Accordingly, we have analyzed the effective mass enhancement as functions of the Coulomb interaction strength and temperature (see Fig.~\ref{fig:meff_nbti}).

In the \textit{bcc}-structure, the Green's function is diagonal in both momentum and spin space and the quasiparticle dispersion can be expressed as: 
\begin{align}
    \tilde{\epsilon}_{(l,m)}^{\sigma}(\mathbf{k},z) = Z_{(l,m)}^{\sigma} \left[ \epsilon_{(l,m)}^{\sigma}(\mathbf{k}) +\Re{\Sigma_{(l,m)}^{\sigma}(z)} \right],
\end{align} 
where $\epsilon_{(l,m)}^{\sigma}(\mathbf{k})$ denotes the bare dispersion.
The quasiparticle density of states is determined by integrating the spectral function over the entire Brillouin zone volume:
\begin{align}
    \tilde{N}_{(l,m)}^{\sigma} (E) = \frac{1}{V_{BZ}} Z_{(l,m)}^{\sigma} \sum_{\mathbf{k}} \delta \left(E - \tilde{\epsilon}_{(l,m)}^{\sigma}(\mathbf{k},E) \right),
\end{align}
such that 
\begin{align}
    N \simeq \sum_{(l,m),{\sigma}} \int_{-\infty}^{E_F} \dd E \, \tilde{N}_{(l,m)}^{\sigma} (E) .
\end{align}

The spectral function seen in Fig.~\ref{fig:nbti_dos_dmft} shows the orbital $(l,m)$ and spin $(\sigma)$ traced and wave vector $(\mathbf{k})$ integrated imaginary part of the Green's function in the non-magnetic state (the self-consistent solution) for different $U$ and $T$ values.  
Despite the locality inherent to DMFT and the perturbative nature of the employed impurity solver, our computations demonstrate that the Luttinger sum rule is satisfied within the limits of numerical accuracy.
This accuracy can be characterized by the difference in chemical potentials (Fermi levels) between the DMFT and LDA calculations, which remains below $\Delta E_F = \tilde{E}_F - E_F^{LDA} \approx \SI{e-5}{\rydberg}$ for all $U$ and at any pressure $P$.
Notably, the enforcement of the Luttinger theorem is ensured by the full self-consistency of the LDA+DMFT framework, which restores translational invariance to the impurity model used in DMFT.

\section{Conclusion}
\label{sec:summary}
We analyzed the behavior of the electronic structure in the normal state of bulk Nb and Nb$_{0.44}$Ti$_{0.56}$ alloys with the $\beta$(\textit{bcc})-structure at pressures up to \SI{250}{\giga\pascal} using a combination of LDA(+CPA) and DMFT methods.
Unlike bulk Nb, where the superconducting critical temperature $T_c$ decreases with pressure, an increase of $T_c$ with pressure in the doped Nb$_{0.44}$Ti$_{0.56}$ alloy has been reported~\cite{st.ti.97}.
The decrease in $T_c$ for pure Nb has been well-documented and is attributed to the position of the Fermi level within a region of low density of states, due to the broadening and spreading of electronic states under pressure~\cite{bo.pa.77,st.ti.97}. Our computations confirm these early findings. 

The remarkable effect of pressure upon the Fermi surface of pure Nb, as described above, is the emergence of a hole band crossing the $(110)$ plane within the first Brillouin zone, which increases in size and becomes distinctly visible at pressures of \SI{150}{\giga\pascal} and above. 
This new band compensates for the reduction in the number of carriers (holes) in the shrinking octahedral shape part of the Fermi surface near the center of the Brillouin zone. The prediction for the appearance of this hole band awaits confirmation by experimental spectroscopic studies.

The inclusion of Ti doping in Nb necessitates a discussion of electronic correlations beyond the standard DFT exchange-correlation functionals. In our study, we employed the regular CPA approach supplemented with the modeling of dynamical electronic correlations in the normal state of the superconducting Nb$_{0.44}$Ti$_{0.56}$ alloy. 
In the disordered alloy with non-magnetic impurities, one might expect the quality of electronic (superconducting) pairing to degrade due to scattering at impurity centers. 
In fact, such pairing must occur between time-reversed exact eigenstates of the metal, even in the presence of defects, as pointed out by Anderson~\cite{ande.59}. 
Another important factor affecting $T_c$ in BCS-Eliashberg-like theories is the cutoff of the phonon frequencies.
Following this argument, several studies have analyzed the pressure effects in Nb$_{0.44}$Ti$_{0.56}$ focusing primarily on phonons, and thus structural supercells with close resemblance to the disordered Nb-Ti \textit{bcc}-structure have been proposed to model these effects.

For example, Refs.~\cite{gu.li.19,zh.ga.20,hu.gu.20} considered a regular CsCl 
structure; more advanced simulations including anharmonic effects were considered in Ref.~\cite{cu.na.24}. 
While these studies provided significant insights into the superconducting properties, none addressed the Fermi surface changes due to pressure.  
Our results fill this gap: we have studied the changes in the total and orbital-resolved density of states, self-energies, and effective mass renormalization due to the presence of interactions and pressure which, to the best of our knowledge, have not been reported previously. In particular, we
demonstrate that both electronic orbitals, t$_{2g}$ and e$_g$, exhibit a significant reaction to pressure, as they both spread and broaden in energy. 
We also conjecture that superconducting pairing will be separable for the different orbitals: pairing may have different strengths, but given the \textit{bcc}-symmetry, no inter-orbital pairing is expected, as the t$_{2g}$ and e$_g$ orbitals are orthogonal, and their Green's functions remain diagonal.

We provide a detailed description and comparison of the Fermi surfaces of pure Nb and Nb$_{0.44}$Ti$_{0.56}$. 
Unlike pure Nb, where the deformation of the jungle-gym shape can lead to a closed orbit, in Nb$_{0.44}$Ti$_{0.56}$ this behavior does not emerge. Nevertheless, the conservation of the hole number is realized through the balance of volumes in various regions of the first Brillouin zone. It is noteworthy that the Luttinger sum rule for the particle densities is excellently fulfilled.
Our computations also show that despite the substantial structural disorder ($56\%$ Ti in Nb), the system retains its Fermi-liquid character, with a moderate mass enhancement up to a factor of $1.5$.
The effective mass decreases with increasing pressure, which suggests that the electronic system becomes less interacting at higher pressures. Consequently, BCS-like theories may be more suitable in the high-pressure regime than at lower pressures.

Finally, we mention possible analogies with the recent theoretical study on the enhanced superconductivity in vanadium-titanium alloys~\cite{jo.os.24} having a similar $\beta$-phase (\textit{bcc}-structure). At an optimal doping of about $33\%$ of Ti  in the host V the superconducting critical temperature is increased, while its normal state is a Fermi liquid with similar characteristics as those discussed here for the Nb-Ti alloy. It is therefore an interesting avenue to investigate the physical properties of V$_{1-x}$Ti$_{x}$ at similarly high pressures in search of a further enhancement of the critical temperatures.   
Our observations (on both Nb-Ti, this paper, and V-Ti alloys~\cite{jo.os.24}) confirm that the physical properties of the normal state are indispensible for describing the mechanism of superconductivity. We demonstrate that even under pressures up to \SI{250}{\giga\pascal}, the normal state does not deviate from the regular Fermi-liquid picture of  Nb-Ti, and well-defined quasiparticles exist along the Fermi surface sheets.

\section*{Acknowledgements}
Financially supported by the Deutsche Forschungsgemeinschaft (DFG, German Research 
Foundation) through TRR 360, project no. 492547816.
LC acknowledges the hospitality of Clarendon Laboratory, Department of Physics, University of Oxford, UK, during the preparation of the manuscript.
LV acknowledges the Swedish Research Council, the Swedish Foundation for Strategic Research, the Carl Trygger Foundation, and the Swedish Foundation for International Cooperation in Research and Higher Education.

\bibliography{sources}

\end{document}